\documentclass[twocolumn]{aastex631}

\usepackage{amsmath,amssymb,amstext}
\usepackage{gensymb}
\usepackage{multirow}
\usepackage{rotating}
\usepackage{tabularx}
\usepackage{comment}
\usepackage[version=4]{mhchem} 
\usepackage{enumitem}
\setlist[itemize]{noitemsep, topsep=0pt, parsep=0pt, partopsep=0pt}

\defcitealias{Teske2025}{Teske \& Batalha et al. 2025}
\defcitealias{May2023}{May \& MacDonald et al. 2023}

\begin{document}

\title{JWST COMPASS: NIRSpec/G395H Transmission Observations of the Sub-Neptune HD 15337~c}

\author[0000-0003-0354-0187]{Nicole L. Wallack}
\affiliation{Earth and Planets Laboratory, Carnegie Institution for Science, 5241 Broad Branch Road, NW, Washington, DC 20015, USA}

\author[0000-0002-8518-9601]{Peter Gao} 
\affiliation{Earth and Planets Laboratory, Carnegie Institution for Science, 5241 Broad Branch Road, NW, Washington, DC 20015, USA}

\author[0000-0002-0371-1647]{Michael Greklek-McKeon}
\affiliation{Division of Geological and Planetary Sciences, California Institute of Technology, Pasadena, CA 91125, USA}

\author[0000-0002-7500-7173]{Annabella Meech}
\affiliation{Center for Astrophysics ${\rm \mid}$ Harvard {\rm \&} Smithsonian, 60 Garden St, Cambridge, MA 02138, USA}

\author[0000-0002-8949-5956]{Artyom Aguichine}
\affiliation{Department of Astronomy and Astrophysics, University of California, Santa Cruz, CA, USA}

\author[0000-0003-4157-832X]{Munazza K. Alam}
\affiliation{Space Telescope Science Institute, 3700 San Martin Drive, Baltimore, MD 21218, USA}

\author[0000-0001-8703-7751]{Lili Alderson}
\affiliation{Department of Astronomy, Cornell University, 122 Sciences Drive, Ithaca, NY 14853, USA}

\author[0000-0003-1240-6844]{Natasha E. Batalha}
\affiliation{NASA Ames Research Center, Moffett Field, CA 94035, USA}

\author[0000-0002-7030-9519]{Natalie M. Batalha}
\affiliation{Department of Astronomy and Astrophysics, University of California, Santa Cruz, CA 95064, USA}

\author[0009-0003-2576-9422]{Anna Gagnebin}
\affiliation{Department of Astronomy and Astrophysics, University of California, Santa Cruz, CA 95064, USA}

\author[0000-0001-5253-1987]{Tyler A. Gordon}
\affiliation{Department of Astronomy and Astrophysics, University of California, Santa Cruz, CA 95064, USA}

\author[0000-0002-4207-6615]{James Kirk} 
\affiliation{Department of Physics, Imperial College London, Prince Consort Road, London SW7 2AZ, UK}

\author[0000-0003-3204-8183]{Mercedes López-Morales}
\affiliation{Space Telescope Science Institute, 3700 San Martin Drive, Baltimore MD 21218, USA}

\author[0000-0002-6721-3284]{Sarah E. Moran}
\affiliation{NASA Goddard Space Flight Center, 8800 Greenbelt Rd, Greenbelt, MD 20771, USA; NHFP Sagan Fellow}

\author[0000-0002-4489-3168]{Jea Iyanla Redai}
\affiliation{Center for Astrophysics ${\rm \mid}$ Harvard {\rm \&} Smithsonian, 60 Garden St, Cambridge, MA 02138, USA}

\author[0000-0003-3623-7280]{Nicholas Scarsdale}
\affiliation{Department of Astronomy and Astrophysics, University of California Santa Cruz, Santa Cruz, CA, 95064, USA}

\author[0009-0008-2801-5040]{Johanna Teske}
\affiliation{Earth and Planets Laboratory, Carnegie Institution for Science, 5241 Broad Branch Road, NW, Washington, DC 20015, USA} 
\affiliation{The Observatories of the Carnegie Institution for Science, 813 Santa Barbara St., Pasadena, CA 91101, USA}

\author[0000-0003-4328-3867]{Hannah R. Wakeford}
\affiliation{School of Physics, HH Wills Physics Laboratory, Tyndall Avenue, Bristol, BS8 1TL}

\author[0000-0002-0413-3308]{Nicholas F. Wogan}
\affiliation{NASA Ames Research Center, Moffett Field, CA 94035, USA}

\author[0000-0003-2862-6278]{Angie Wolfgang}
\affiliation{Eureka Scientific}

\begin{abstract}
We present the 3-5 $\mu$m transmission spectrum of HD 15337~c (TOI-402.02), a sub-Neptune (2.526 R$_{\earth}$, 6.792 M$_{\earth}$, T$_{\rm eq}$$\sim$656 K) around a K1V star observed as part of the JWST COMPASS program. We reduce these observations with two pipelines and find consistent transmission spectra. The resulting median precisions in 30 pixel spectroscopic bins for visit 1 are $\sim$40 ppm and $\sim$70 ppm and for visit 2 are $\sim$30 ppm and $\sim$54 ppm for NRS1 and NRS2, respectively. We attribute the differing precisions to the lack of adequate pre-transit baseline in visit 1 from an early transit arrival caused by previously undetected transit timing variations (TTVs), hinting at a potential exterior companion. Our median JWST timing precision is 10 seconds, revealing TTVs $>$20 minutes when combined with previous TESS and CHEOPS data, highlighting JWST's TTV measurement capabilities. The transmission spectrum of HD 15337~c is featureless and can best be described by a step function with an offset between the NRS1 and NRS2 detectors, likely caused by instrumental systematics. From thermochemical equilibrium retrievals we find that, to $>$3$\sigma$, the data can rule out atmospheres with metallicities $<$600 or $<$310 $\times$ solar, depending on the reduction, for opaque pressures greater than a few millibars. HD 15337~c joins other sub-Neptunes with similar masses, radii, and temperatures in possessing a featureless transmission spectrum indicative of high metallicity and/or high-altitude aerosols and adds support to recent studies showing that aerosol opacity reaches a maximum for planets with equilibrium temperatures of 500-700 K. 

\end{abstract}

\section{Introduction} 
\label{sec:intro}

Sub-Neptunes are the most common type of planet in the Galaxy discovered to date \citep{Batalha2013, Fulton2017}, and yet their formation and evolution have been hotly debated. These planets (1.7 R$_\earth$ $\lesssim$ R$_{\rm p}$ $\lesssim$ 4 R$_\earth$), which are on the larger side of the radius valley from their smaller super-Earth counterparts (R$_{\rm p}$ $\lesssim$ 1.7 R$_\earth$), are thought to perhaps be the scaled-down versions of gas giant planets, with primordial hydrogen-helium envelopes  \citep[e.g.,][and references therein]{Bean2021}. However, our understanding of the formation, evolution, and structure of these sub-Neptunes is still evolving. Possible formation theories for these sub-Neptunes include formation outside of the water ice-line with migration to their current orbits or initial formation in gas-rich environments versus the gas-poor environments of super-Earths \citep[e.g.,][]{Lee2022, Raymond2018, Burn2021, Burn2024}. The different formation mechanisms of these planets are thought to result in differing observable atmospheres, making characterization of their atmospheres crucial for our understanding of their formation histories. 

However, it is not only the formation location that can impact a planet's observable atmosphere. Planetary system parameters, such as the temperature of the planet and host star properties, can impact the evolution of a planet's atmosphere and therefore what we can infer about the present-day state of the planet. Despite the unparalleled sensitivity and precision allowed by JWST, recent JWST transmission observations of sub-Neptunes have not been entirely straightforward to interpret from a population standpoint. To date, the only JWST transmission observations of larger sub-Neptunes (2.4 R$_{\oplus}$ $<$ R$_{\rm p}$$<$4 R$_{\oplus}$) that have been published are GJ 1214~b \citep{Schlawin2024, Ohno2025}, K2-18~b \citep{Madhu2023, Schmidt2025}, TOI-836~c \citep{Wallack2024}, and TOI-421~b \citep{Davenport2025}. While we might expect these larger sub-Neptunes to be more compositionally homogeneous than their smaller sub-Neptune counterparts that are closer to the radius valley and which may have more diversity \citep{Piaulet-Ghorayeb2024}, the results of these studies have indicated that these sub-Neptunes also have diverse atmospheres with a range in atmospheric compositions and cloud or aerosol coverage. The warmer objects (500 $<$ T$_{\rm eq}$ $<$ 800 K) seem to have high altitude aerosols that challenge our ability to constrain their atmospheric compositions, while the coolest and hottest planets appear to be devoid of aerosols, which seems to be consistent with the picture inferred from the sizes of water features at 1.4 $\mu$m from Hubble Space Telescope (HST) observations \citep[e.g.,][]{Brande2024, Dymont2022}. The picture becomes even more complex when including the smaller sub-Neptunes (R$_{\rm p}$$<$ 2.4 R$_{\oplus}$), GJ 9827~d \citep{Piaulet-Ghorayeb2024}, TOI-270~d \citep{Benneke2024}, TOI-776~c (\citetalias{Teske2025}), and GJ 3090~b \citep{Ahrer2025}, where planets of the same size and temperature have very different spectra (e.g. TOI-270~d versus TOI-776~c and GJ 3090~b versus GJ 9827~d). For example, while TOI-270~d (2.2 R$_{\oplus}$, 4.78 M$_{\oplus}$,$\sim$380 K; \citealt{Benneke2024}) and TOI-776~c (2.047 R$_{\oplus}$, 6.9 M$_{\oplus}$,$\sim$420 K; \citetalias{Teske2025}) have comparable bulk compositions and temperatures, TOI-270~d shows evidence for the presence of water, carbon dioxide, and methane \citep{Benneke2024}, in contrast to TOI-776~c, which has a featureless transmission spectrum consistent with an atmosphere of high mean molecular weight and/or the presence of high-altitude aerosols (\citetalias{Teske2025}). Similarly, while  GJ 3090~b (2.13 R$_{\oplus}$, 3.34 M$_{\oplus}$, $\sim$700 K; \citealt{Ahrer2025}) and GJ 9827~d (1.98 R$_{\oplus}$, 3.02 M$_{\oplus}$,$\sim$620 K; \citealt{Piaulet-Ghorayeb2024}) are also quite comparable in their bulk parameters, GJ 9827~d shows evidence of having a water-rich atmosphere with an absence of aerosols \citep{Piaulet-Ghorayeb2024}, whereas the transmission spectrum of GJ 3090~b is generally featureless aside from evidence of metastable helium absorption, consistent with an atmosphere of high mean molecular weight \citep{Ahrer2025}.  
The wide range in atmospheric compositions of sub-Neptunes observed with JWST underpins the need for more observations in order to understand this population and determine whether suggested properties such as equilibrium temperature \citep[e.g.,][]{Brande2024} do indeed drive the presence of aerosols on sub-Neptunes.

In order to better understand the formation and evolution of the population of small planets more generally, we initiated the JWST COMPASS (Compositions of Mini-Planet Atmospheres for Statistical Study) Program. COMPASS is a JWST Cycle 1 Program \citep[PID \# 2512;][]{2021jwst.prop.2512B}, which is looking at a population of 12 super-Earths and sub-Neptunes in transmission with NIRSpec/G395H (including 11 new observations), in order to statistically understand this population. The COMPASS targets are a specifically selected sub-population of the Magellan-TESS Survey (MTS, see \citealt{2021ApJS..256...33T}), where targets were chosen from the Year 1 TESS targets using a quantitative ranking metric that is a function of observing time required for photon-limited radial velocity (RV) precision, stellar insolation, and planet radius. The down-selection of targets from MTS for COMPASS were quantitatively selected via a metric that further weighed the insolation flux, stellar effective temperature, planet radius, and the exposure time required to obtain 30~ppm precision at 4~$\mu$m using NIRSpec/G395H binned to R=100 \citep[see][for more details]{Batalha2023}. Taken together, the COMPASS sample was explicitly selected to be representative of the population of super-Earth and sub-Neptune planets in order to allow for population-level trends to be determined from the analyses of individual planetary spectra (\citealt{ Alderson2024,Scarsdale2024, Wallack2024, Alam2025, Alderson2025, Adams2025}; \citetalias{Teske2025}). 

Here we present transmission spectra from two NIRSpec/G395H observations of HD 15337~c (TOI-402.02), the last of three sub-Neptunes observed as part of the COMPASS program. The multi-planet HD 15337 system was simultaneously discovered by both \cite{Dumusque2019} and \cite{Gandolfi2019} and consists of two planets of similar masses but differing radii. HD 15337~c is a 6.792 $^{+1.302}_{-1.143}$  M$_{\earth}$, 2.526 $^{+0.086}_{-0.075}$ R$_{\earth}$, 2.303$^{+0.505}_{-0.414}$ g/cm$^{3}$ planet on an orbit exterior to the smaller HD 15337~b \citep[6.519  $^{+0.409}_{-0.400}$ M$_{\earth}$, 1.770  $^{+0.032}_{-0.030}$ R$_{\earth}$, 6.458$^{+0.518}_{-0.510}$ g/cm$^{3}$;][]{Rosario2024}. HD 15337~c, and the architecture of the HD 15337 system (Table~\ref{table:system}), is markedly similar to the other large sub-Neptune observed as part of the COMPASS program, TOI-836~c \citep{Wallack2024}, which also orbits a K star, albeit one of a later type (T$_{\rm eff}$= 4552 K, \citealt{Hawthorn2022}).

We describe our observations in Section \ref{sec:obs}, and we detail our data reduction methods and light curve fitting methods in Section \ref{sec:data_reduction}. We present updated orbital parameters for HD 15337~c in Section \ref{sec:orbit} and its transmission spectrum in Section \ref{sec:spectrum}, and interpret our results using nonphysical and thermochemical equilibrium atmospheric models in Section \ref{sec:modeling}. We put our results in the context of other sub-Neptunes observed with JWST in Section \ref{sec:discussion} before summarizing our results in Section \ref{sec:summary}.  

\begin{deluxetable}{ll}

\tablewidth{0pt}
\tablehead{\colhead{Property}&\colhead{Value}}
\startdata 
G (mag)  & 8.865 $\pm$0.00276 \\
T$_*$ (K)& 5131$\pm$74\\
R$_*$ (R$_\odot$)&  0.855$\pm$0.008\\
log(g) & 4.37$\pm$0.08\\
$[$Fe/H$]_{*}$& 0.03$\pm$0.04\\
\hline
Period (days)& 17.180546 $^{+0.000021}_{-0.000026}$\\
Mass (M$_{\earth}$)&  6.792 $^{+1.302}_{-1.143}$ \\
Radius (R$_{\earth}$)& 2.526  $^{+0.086}_{-0.075}$\\
T$_{\tt eq}$ (K)& 656.1 $^{+8.4}_{-9.0}$\\
$e$ & 0.096$^{+0.059}_{-0.045}$\\
$\omega$ & 56.63$^{+77.89}_{-63.30}$\\
inclination (\textdegree)&88.41 $\pm$ 0.07\\
a (AU)& 0.1235$^{+0.0020}_{-0.0017}$\\
a/R$_*$& 31.63$^{+0.82}_{-0.77}$
\tablecaption{System Properties for HD 15337~c}
\enddata 
\label{table:system}
\tablenotetext{}{All values from \cite{Rosario2024}, with the exception of a, a/R$_*$, and T$_{\tt eq}$, which are not reported by \cite{Rosario2024} and are from \cite{Dumusque2019}.}
\end{deluxetable}

\section{Observations} 
\label{sec:obs}

\subsection{NIRSpec}

We observed two transits of HD 15337~c with JWST/NIRSpec on 2023 July 24 and 2023 August 27 using the high-resolution (R$\sim$2700) G395H mode, which provides spectroscopy between 2.87--5.14\,$\mu$m across the NRS1 and NRS2 detectors (with a gap between 3.72--3.82\,$\mu$m). The observations were taken with the NIRSpec Bright Object Time Series (BOTS) mode using the SUB2048 subarray, the F290LP filter, the S1600A1 slit, and the NRSRAPID readout pattern.  Each $\sim$5.86 hour exposure consisted of 4654 integrations (divided into four segments) with four groups per integration, and was designed to be centered on the $\sim$2.2 hour transit event with sufficient out-of-transit baseline. Owing to the large uncertainty in the ephemeris using the \cite{Dumusque2019} and \cite{Gandolfi2019} values (which used the three TESS transits of HD 15337~c observed at the time of their publication), we opted to use the updated ephemeris available on ExoFOP which utilized all five TESS transits available at the time of scheduling the JWST observations.

\begin{figure}
    \centering
    \includegraphics[width=.45\textwidth]{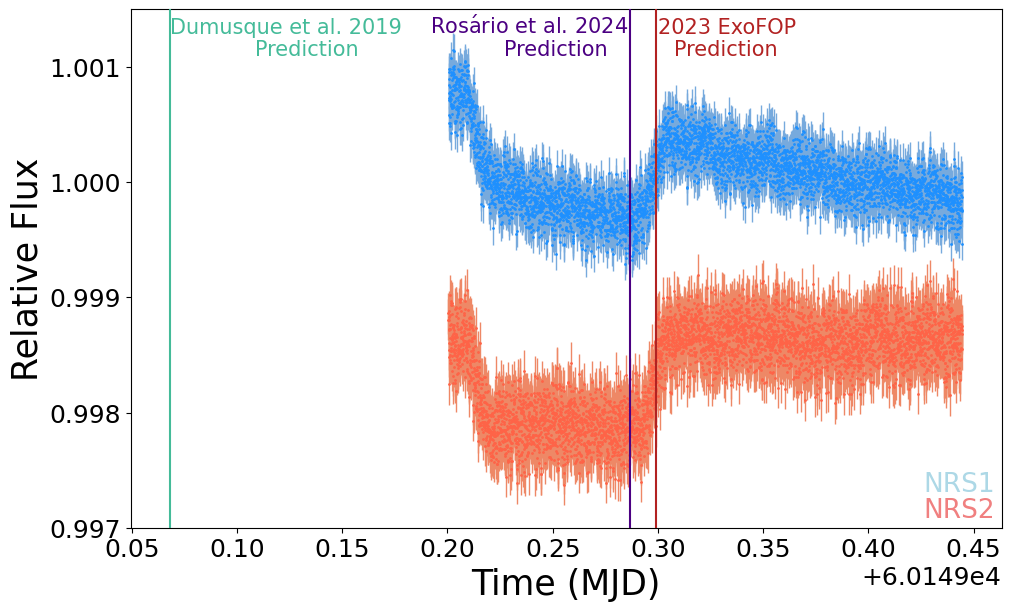}
    \caption{The white light curves from visit 1 of HD 15337~c with the predicted midtransit times from \cite{Rosario2024} in purple, \cite{Dumusque2019} in green, and from ExoFOP in red. The transit of HD 15337~c occurred $\sim$44 minutes earlier than the time predicted from \cite{Rosario2024}, $\sim$61 minutes earlier than the time predicted from ExoFOP, and $\sim$4.5 hours later than the time predicted from \cite{Dumusque2019}.}
    \label{fig:Visit1_WLC}
\end{figure}

However, upon initial inspection of the first of our two observations, it was evident that the transit of the planet was not well centered, indicating the potential for a stale ephemeris and/or period (Figure~\ref{fig:Visit1_WLC}), a transit timing variation (TTV), or both. We reduced and fit the data of this first transit in the manner described below, finding that it deviated by $\sim$4.5 hours from the time predicted using the best-fit values from \cite{Dumusque2019} and $\sim$61 minutes from the updated ephemeris available on ExoFOP. We then centered the subsequent transit (which occurred two planetary orbits later) using the best-fit transit time from the first observation, and found that this resulted in the transit being well centered within our observation (Figure~\ref{fig:WLCs}). 

\section{NIRS\MakeLowercase{pec}/G395H Data Reduction} 
\label{sec:data_reduction}

\subsection{Eureka!}
We first used the \texttt{Eureka!}\ pipeline to reduce our observations. \texttt{Eureka!}\ is an end-to-end pipeline for the analysis of JWST and HST observations \citep{Bell2022}. We utilize \texttt{Eureka!}\ in a similar manner to that described in previous COMPASS papers \citep[e.g.,][]{Wallack2024}, and briefly summarize that process herein. The \texttt{Eureka!}\ JWST pipeline acts as a wrapper around the \texttt{jwst} pipeline for the first two calibration stages with the addition of a custom group level background subtraction correction. We use \texttt{Eureka!}\ version 0.10 and \texttt{jwst} version 1.11.4 with context map jwst\_1293.pmap. We use all of the default Stage 1 and Stage 2 \texttt{jwst} stages (with the inclusion of the \texttt{Eureka!}\ group level background subtraction), with the exception of utilizing a jump detection threshold of 15$\sigma$. 

We then utilize \texttt{Eureka!}\ Stage 3 and Stage 4, which do additional calibration on the data as well as extract the time series stellar spectra. As Stage 3 of \texttt{Eureka!}\ has a number of different reduction-specific choices in its further calibration steps, we ascertain the preferred reduction parameters for our observations by optimizing our reduction in a similar manner to that described in \cite{Wallack2024}. We select the extraction aperture, background aperture, polynomial order for an additional background subtraction, and sigma threshold for the outlier rejection during the optimal extraction. We generate white light curves (summing over 2.863--3.714 microns for NRS1 and 3.820--5.082 microns for NRS2) and 30 pixel binned spectroscopic light curves for multiple versions of the reduction assuming extraction apertures of between 4 and 8 pixel half-widths, background apertures of between 8 and 11 pixels (measured from the edge of the trace to the edge of the frame in both directions), additional column-by-column or full frame background subtraction, and sigma thresholds for the outlier rejection during the  optimal extraction of 10 and 60. We generate light curves for all combinations of these parameters and select the combination of parameters that minimizes the median absolute deviations across all of the spectroscopic light curves. We optimize each of these parameters for each observation and detector separately and find preferred combinations of 4-pixel extraction apertures and 8-pixel background apertures for both visits and both detectors. We also find that an additional linear column-by-column background subtraction is preferred for both detectors and both visits, and an optimal extraction outlier rejection sigma threshold of 60 is preferred over 10 for all but visit 2 of NRS2.

Instead of using the light curve fitting code built into Stage 5 and Stage 6 of \texttt{Eureka!}, we opt for a custom light curve fitting code for increased flexibility as we have done in previous COMPASS papers (e.g., \citealt{Alderson2024, Wallack2024}). We use both the inbuilt \texttt{Eureka!} outlier rejection and a custom outlier rejection for the white light curves and spectroscopic light curves from Stage 4 of \texttt{Eureka!}. We first trim 10 sigma outliers from a 50 point rolling median five times using the inbuilt \texttt{Eureka!} outlier rejection, then we also iteratively trim three sigma outliers from a 50 point rolling median three times to remove additional outliers in the time series prior to our custom light curve fitting. This second outlier rejection using a lower sigma threshold does not favor the removal of more than ten integrations for any of our light curves. 

We first fit the white light curves using a Levenberg-Marquardt minimization where we minimize the log likelihood, fitting for the inclination ($i$), the ratio of the semi-major axis to the stellar radius ($a/R_{*}$), the time of transit ($T_{0}$), and the ratio of the radius of the planet to the stellar radius ($R_{p}/R_{*}$) using \texttt{batman} \citep{Kreidberg2015}. We fix the quadratic limb-darkening coefficients to the theoretical values computed with Set One of the MPS-ATLAS models using the stellar parameters in Table~\ref{table:system} with {\tt ExoTiC-LD} \citep{Grant2024}. We simultaneously fit  an instrumental noise model of the form 
\begin{equation}
S= p_{1} + p_{2}\times T+ p_{3}\times X + p_{4}\times Y , 
\label{eq:1}
\end{equation}
where $p_{N}$ is a parameter fitted for in our instrumental noise model, $T$ is the array of times, and $X$ and $Y$ are arrays of the positions of the trace, along with an additional per point error added in quadrature to the measurement errors. We also fix the eccentricity and longitude of periastron, $\omega$, to values in Table~\ref{table:system}. We then initialize 3$\times$ the number of free parameters as walkers using the Markov chain Monte Carlo (MCMC) package \texttt{emcee} \citep{Foreman-Mackey2013} to the best-fit parameters from the Levenberg-Marquardt minimization, fitting for the same free parameters. For all fits, we first run 50,000 steps which are discarded followed by a 50,000-step production run.  

NIRSpec observations can have small initial ramps at the beginning of the observations, which can bias the systematic noise model (e.g., \citealt{Alderson2024, Wallack2024,Schlawin2024}; \citetalias{May2023}). In order to determine the optimal number of integrations to discard, we run the aforementioned MCMC fit for different versions of each of the two white light curves for visit 2, where we trim the initial 0--1000 points in steps of 100. We do not remove any data from the beginning of visit 1 for either detector due to the aforementioned lack of pre-transit baseline. For visit 2, we select the amount of initial data to trim for each detector that best minimizes the difference between the expected white noise scaling and the actual noise in the data with bin size, optimizing each detector separately. We find an optimal trim duration of 900 points ($\sim$68 minutes) for NRS1 and 700 points ($\sim$53 minutes) for NRS2, and adopt these versions of the reductions for our final fits. 

\begin{figure*}
    \centering
    \includegraphics[width=\textwidth]{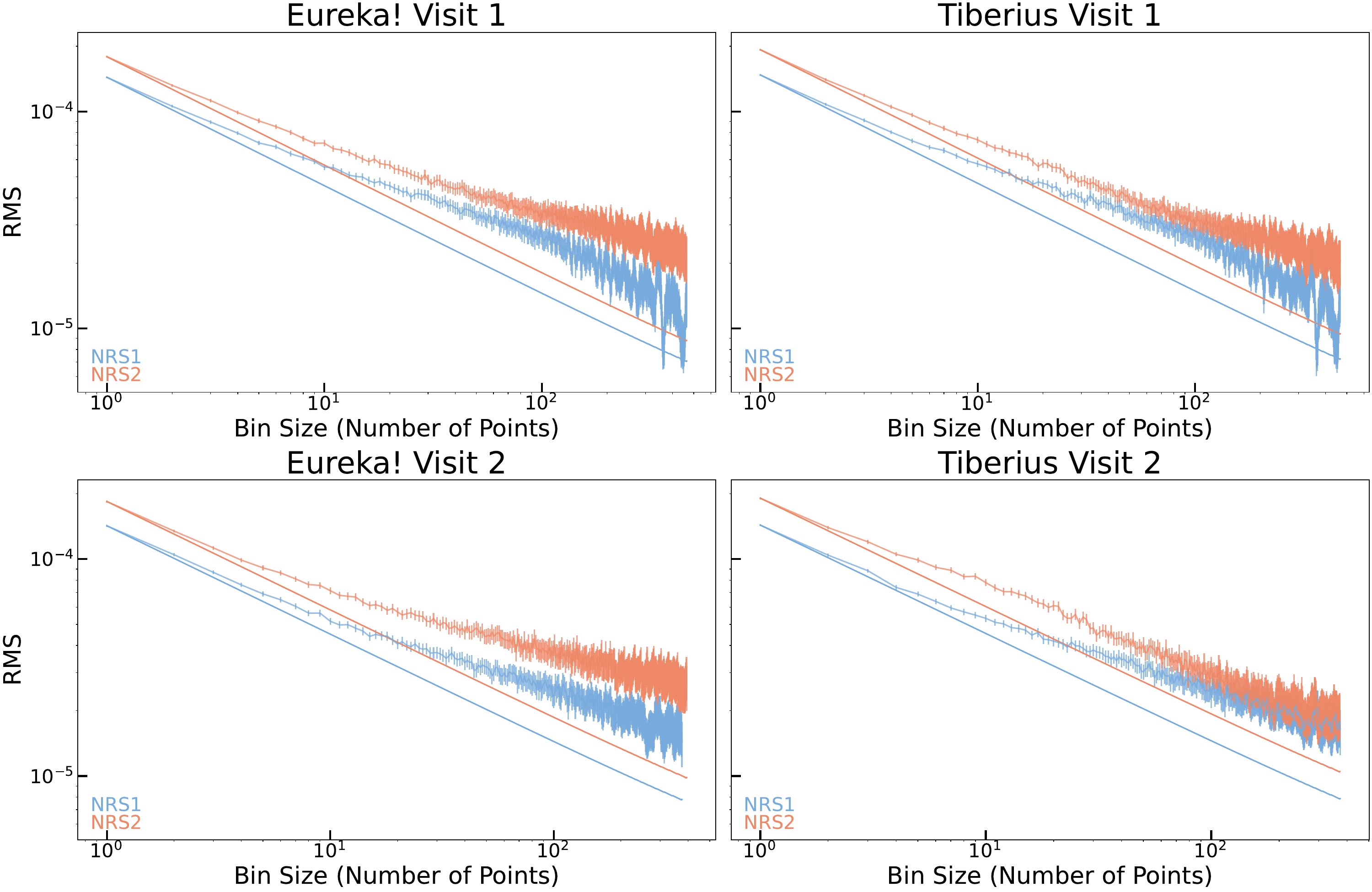}
    \caption{The plots of RMS versus bin size for both detectors for both reductions are shown. We show the actual noise for NRS1 in blue and for NRS2 in red; in the absence of red noise, the residuals would follow the correspondingly colored solid lines. We see residual correlated noise in both detectors for both reductions of both visits.}
    \label{fig:RMS}
\end{figure*}

For our final adopted fits, we first consider each detector and visit separately, assuming different astrophysical and instrumental parameters, in order to determine the presence of any offsets in these parameters. We then do a joint fit to both visits, linking the astrophysical parameters for different visits of the same detector for both the white light curves as well as the spectroscopic light curves. For the joint fits, we do not link the time of transit or the noise model and error inflation terms; these are still considered independent across visits. We further discuss our updated astrophysical white light curve parameters in Section~\ref{sec:orbit}. We show our white light curves, best fit models, and associated residuals from the individual fits in the top panels in Figure~\ref{fig:WLCs} and the associated plots of RMS versus bin size for these white light curves in Figure~\ref{fig:RMS}. 

\begin{figure}
\begin{centering}
\includegraphics[width=0.49\textwidth]{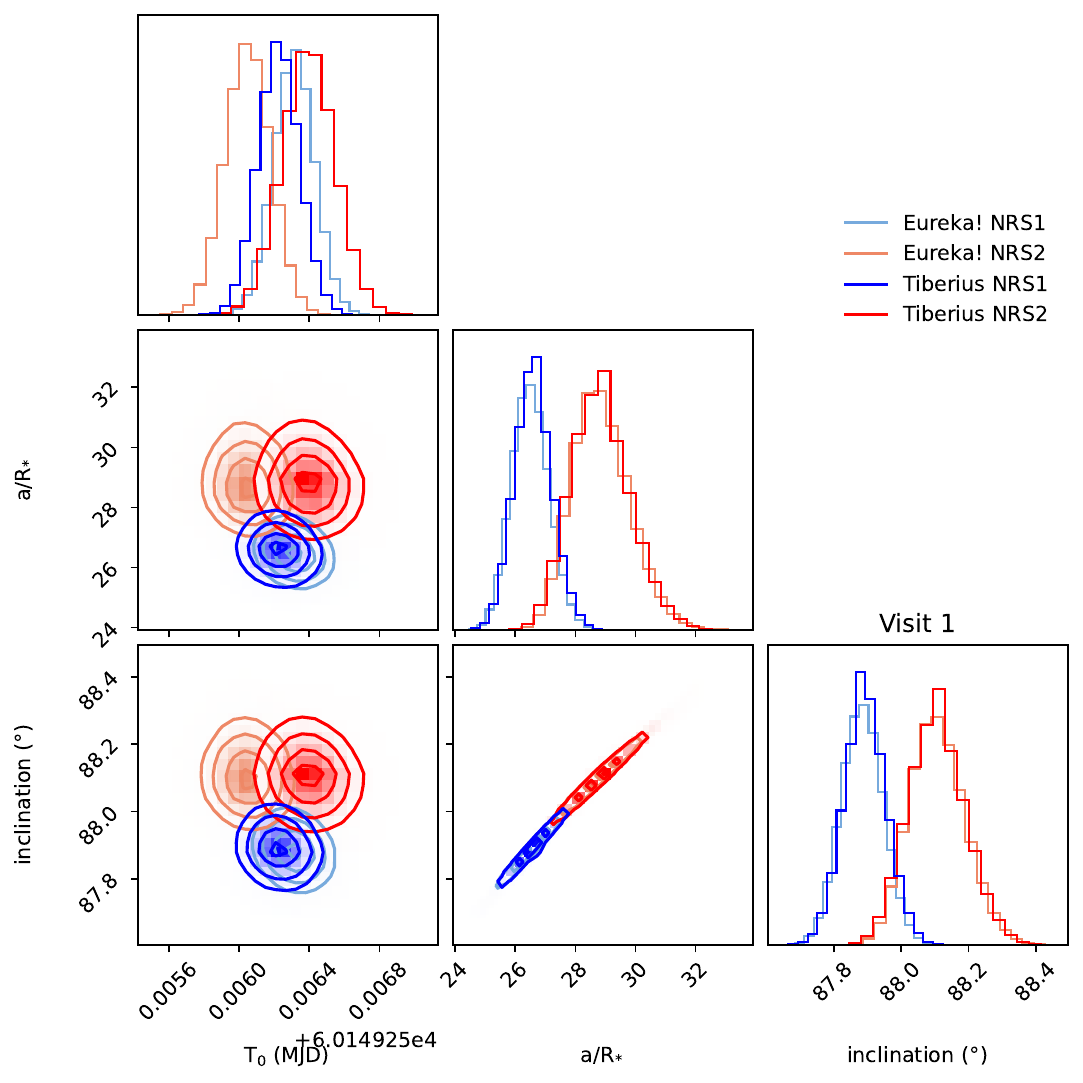} \includegraphics[width=0.49\textwidth]{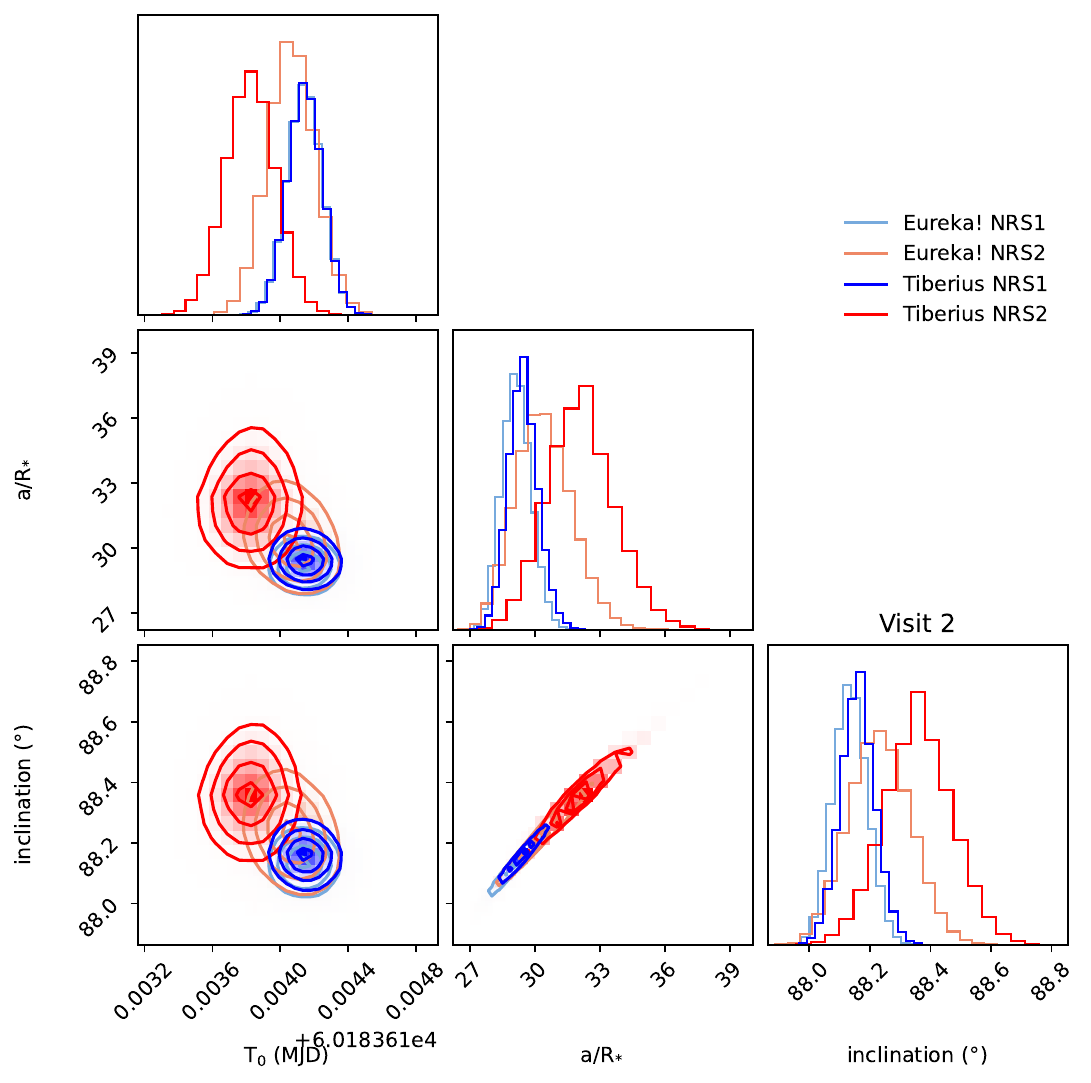} 
\caption{Results from the white light curve fits for visit 1 (top) and visit 2 (bottom) for the \texttt{Eureka!}\ and \texttt{Tiberius} reductions. There is good agreement between the reductions.}
\label{fig:Corner}   
\end{centering}
\end{figure}

\begin{figure}
\begin{centering}
\includegraphics[width=0.45\textwidth]{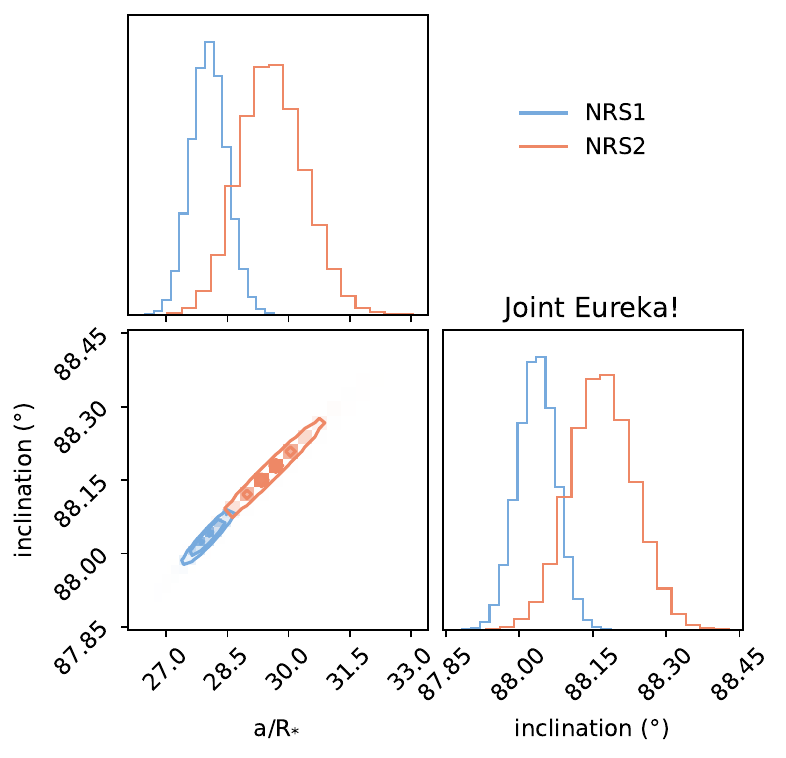} 
\caption{Best fit parameters from the joint fits to the \texttt{Eureka!}\  white light curves. There is good agreement between the joint fit and the individual fits (Figure~\ref{fig:Corner}).}
\label{fig:Corner_Joint}   
\end{centering}
\end{figure}

\begin{figure*}
    \centering
    \includegraphics[width=\textwidth]{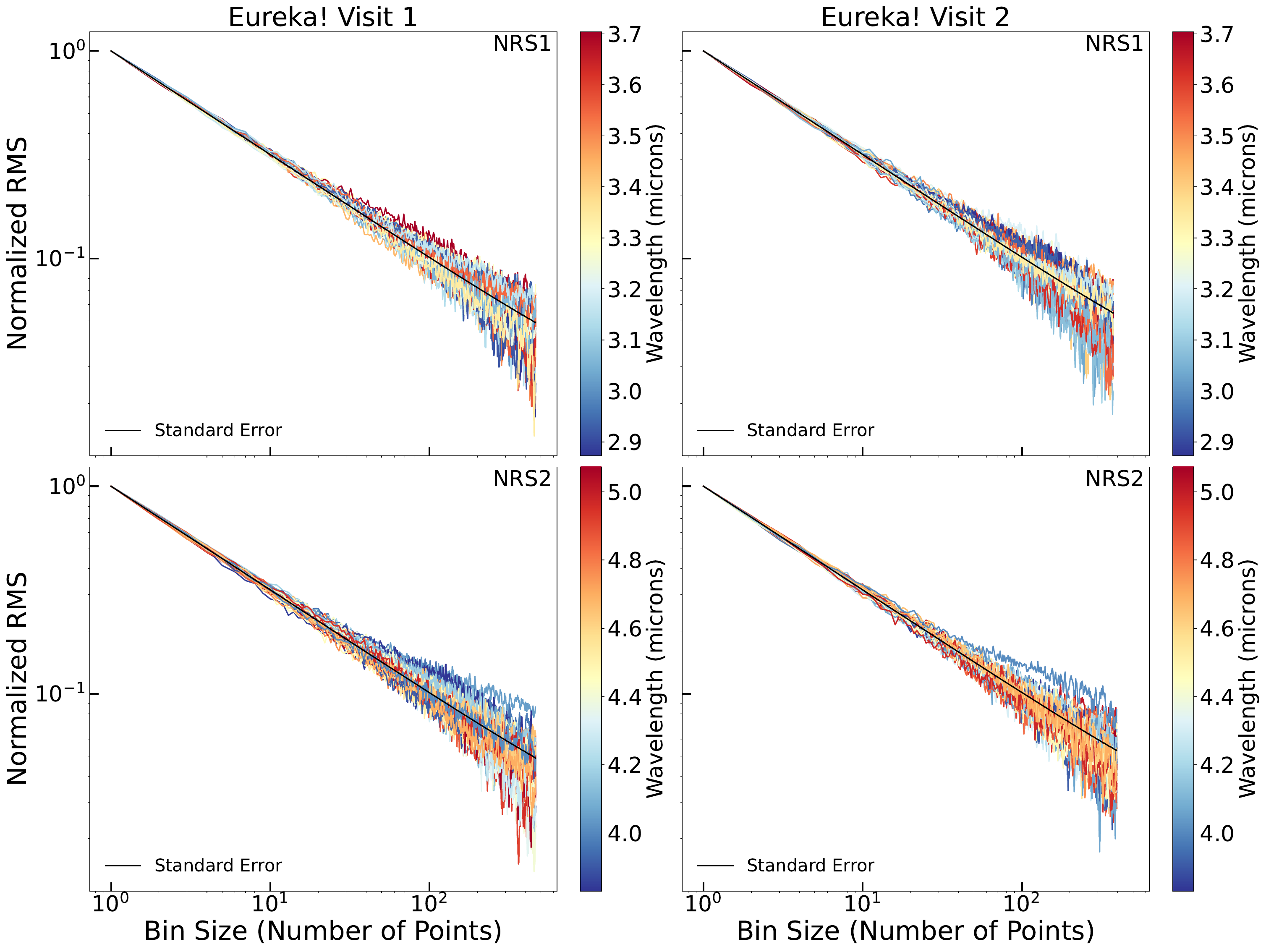}
    \caption{The normalized root mean squared versus bin size plots for both detectors for both visits are shown for the \texttt{Eureka!}\ reduction colored by wavelength. The bins generally follow the black lines, which represent the  standard error, and show that the residual noise seen in the white light curves (Figure ~\ref{fig:RMS}), is not as evident in the individual spectroscopic bins, likely due to the larger noise in the spectroscopic bins.}
    \label{fig:RMS_bins}
\end{figure*}

We fit the 30 pixel binned spectroscopic light curves in the same manner described above, initializing 3$\times$ the number of free parameters for the MCMC walkers at the best-fit of a Levenberg-Marquardt minimization. We fix the $i$, $a/R_{*}$, and $T_{0}$ for all spectroscopic bins to the best fit white light curve parameters within a detector (Table~\ref{tab:fitted-values} and Figures~\ref{fig:Corner} and \ref{fig:Corner_Joint}), leaving  the $R_{p}/R_{*}$ as the only free astrophysical parameter, and fitting a systematic noise model and the additional error term for each spectroscopic bin (see Figure~\ref{fig:RMS_bins} for the associated \texttt{Eureka!}\ plots of RMS versus bin size for each spectroscopic bin). We present our transmission spectra from each individual visit, our jointly fit transmission spectrum, and the error-weighted average of the two visits for comparison to our jointly fit spectrum in Section ~\ref{sec:spectrum}.

\subsection{Tiberius}\label{sec:tiberius}
We used the \texttt{Tiberius} pipeline \citep{Kirk2017,Kirk2021} to provide an independent reduction of the NIRSpec/G395H data. \texttt{Tiberius} uses the STScI \texttt{jwst} pipeline (v1.13.4) to process the detector images. We ran the Stage 1 processing on the uncalibrated fits files (\texttt{uncal.fits}) and performed a custom $1/f$ correction at the group level. Toward this background subtraction, we create a master (median) \texttt{uncal} detector image, and fit the aperture trace with a 4th order polynomial. Taking this trace as the center of the aperture, we then define an aperture width (8 pixels) and background offset (4 pixels) to create an aperture ``mask.'' Having masked the aperture, we compute the median count level in each pixel column, and then subtract this from the unmasked detector image. The final steps of the Stage 1 processing include the ramp fitting and gain scale step, wherein the groups are combined.

\begin{figure*}
    \centering
    \includegraphics[width=\textwidth]{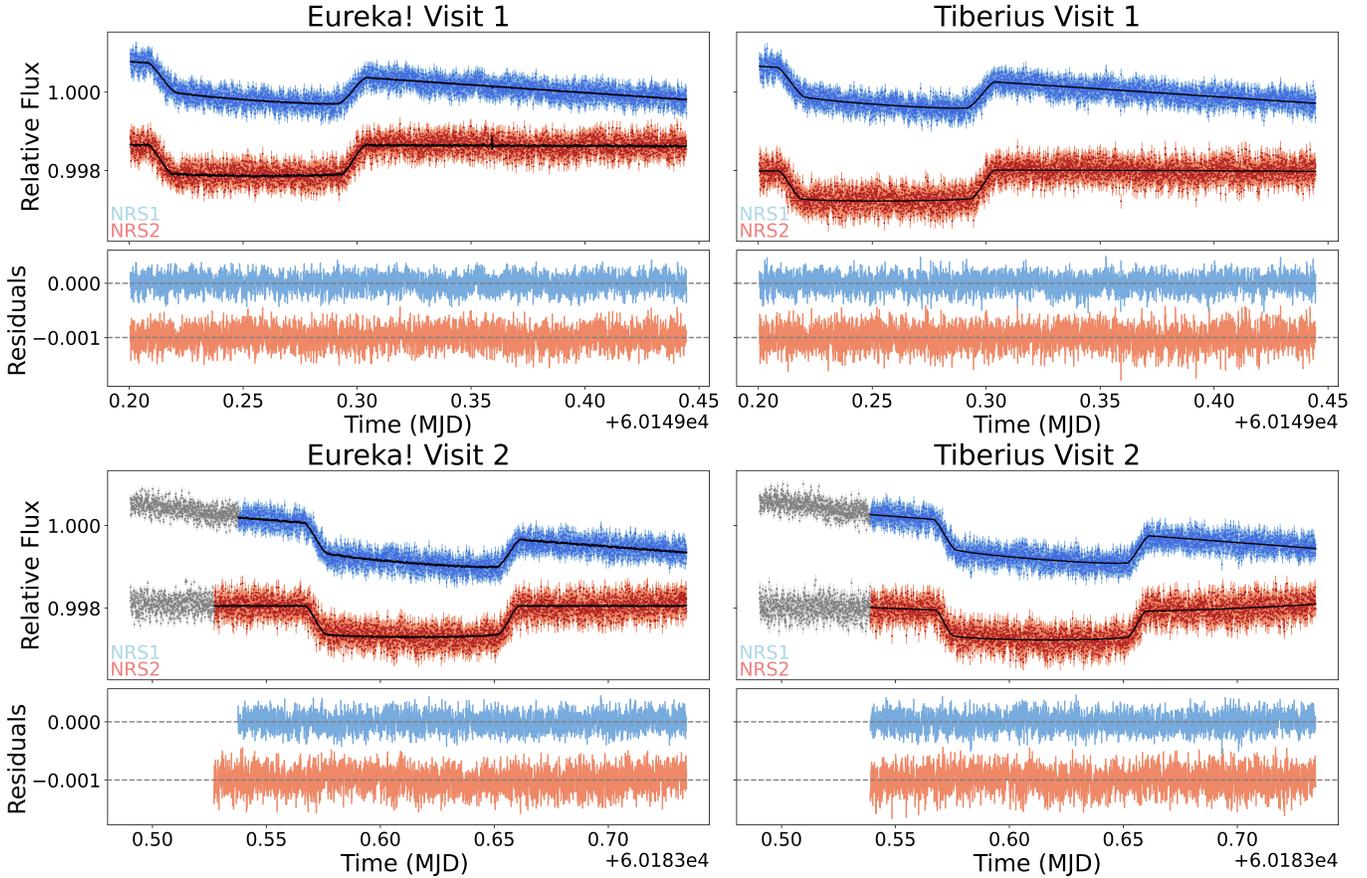}
    \caption{White light curves and residuals from the \texttt{Eureka!}\ (left column) and \texttt{Tiberius} (right column) reductions for visit 1 (top row) and visit 2 (bottom row) of HD 15337~c (in blue for NRS1 and red for NRS2). The gray points in the bottom panels are the trimmed data points from the second visit of the \texttt{Eureka!} and \texttt{Tiberius} reductions, which do not have inflated errors and are not considered in the fitting.}
    \label{fig:WLCs}
\end{figure*}

We proceed with the \texttt{gainscalestep.fits} files; first we run a check for outliers due to e.g., cosmic rays. Assessing the running median counts across the time series, we flag any pixel with a value greater than $4\sigma$ from the median. These and the \textsc{do\_not\_use} pixels, flagged during the \texttt{dq\_init} step, are interpolated over by taking the median of the surrounding pixels. We then extract the stellar spectra: we fit the aperture trace as before and sum the stellar flux in each pixel column within the defined aperture, between pixels $[600, 2040]$ for NRS1 and $[5, 2040]$ for NRS2. Having integrated over all wavelengths, the resulting white light curves are shown in Figure~\ref{fig:WLCs}. We also produce spectroscopic light curves, integrating the stellar spectrum over spectroscopic bins with widths of 30 pixels.

For the light curve fitting, we employ a joint systematics $+$ transit light curve model, with the latter instantiated using the \texttt{batman} package \citep{Kreidberg2015}. Similarly to the \texttt{Eureka!} reduction, we trim the first 70\,minutes (924 integrations) of the visit 2 light curves to remove the initial ramp at the beginning of the observation, to aid in better fitting of the light curves. We do not perform any trimming on the visit 1 light curves. Fitting all four white light curves individually, we explore the parameter space using \texttt{emcee} \citep{Foreman-Mackey2013}, with 30 walkers per parameter, and $10,000$ burn-in followed by $10,000$ production steps. Regarding the astrophysical parameters, we hold the period and limb-darkening coefficients fixed, and leave $a/R_*$, $i$, $T_0$ and $R_p/R_*$ as free parameters. For the limb darkening, we employ a quadratic limb-darkening law. Using the \texttt{ExoTiC-LD} package, we retrieve the coefficients from the 3D \texttt{STAGGER} grid for the stellar parameters given in Table~\ref{table:system}, \citep[namely $u_1=0.0949$, $u_2=0.1732$ for NRS1; $u_1=0.0836$, $u_2=0.1322$ for NRS2;][]{Grant2024}. We fix the eccentricity and the argument of periastron to the values given in Table~\ref{table:system}. We trialed a number of systematics models to detrend the light curves, opting for a linear combination of polynomials, based on the Bayesian Information Criterion (BIC) and lowest residual red noise. A linear polynomial in time was sufficient to detrend the NRS1 white light curves, and we used quadratic time polynomials for the NRS2 white light curves, of the form
\begin{align}
S_\mathrm{NRS1}&= p_{1} + p_{2}\times T; \\
S_\mathrm{NRS2}&= p_{1} + p_{2}\times T + p_{3}\times T^2
\label{eq:tiberius-WLC}
\end{align}
respectively. The white light curves and their corresponding models are shown in Figure~\ref{fig:WLCs}, with the parameters given in Table~\ref{tab:fitted-values}. With \texttt{Tiberius}, we achieve precisions of 147\,ppm and 143\,ppm for the visit 1 and 2 NRS1 light curves, and 191\,ppm for both NRS2 light curves.

To produce a transmission spectrum, we fit the depths of the 30 pixel bin transit light curves.
Along with the period, we fix $a/R_*$, $i$, and $T_0$ to the fitted white light curves values (Table~\ref{tab:fitted-values}). Once again, we use a quadratic limb-darkening law; for each spectroscopic bin, we retrieve the 3D \texttt{STAGGER} limb-darkening coefficients corresponding to the stellar parameters in Table~\ref{table:system}, and leave these fixed. For the systematics model, we fit a linear trend to each spectroscopic light curve. We therefore have three free parameters: $R_\mathrm{p}/R_*$, and the two linear trend coefficients. In Section ~\ref{sec:spectrum}, we present the resulting transmission spectra from each visit and the composite transmission spectrum of HD 15337 c determined by taking the weighted average of the two visits.

We note that our systematic noise model differs between our NRS2 white light curves and NRS2 spectroscopic light curves. We do not find that a higher order time term is preferred by the data for the spectroscopic bins, but we find that it is preferred for the white light curves. Using a quadratic versus linear-only time term to detrend the visit 2 NRS2 white light curve results in astrophysical best-fit parameters that differ by only $<$0.7$\sigma$. The exception is $R_\mathrm{p}/R_*$, which differs by $\sim$2.7$\sigma$ between the two systematic noise models. However, the resulting spectra (having fixed $a/R_*$, $i$ and $T_0$ according to the different white light curve systematics models) differ by less than 0.21$\sigma$ (14 ppm) per spectral bin, as we always utilize a linear time term for our spectroscopic fits. We find that all of the best-fit astrophysical parameters for visit 1 of NRS2 agree within 1$\sigma$, including the $R_\mathrm{p}/R_*$. 

\begin{table*}[]
    \centering
    \caption{Fitted Parameters from the \texttt{Eureka!}\ and \texttt{Tiberius} White Light Curves}
    \begin{tabular}{c c c c c c c}
    \hline
    \hline
         Pipeline& Visit & Detector & $R_p/R_*$ & $T_0$\,[MJD] & $i$\,[$^\circ$] & $a/R_*$  \\
         \hline
         Eureka! & Visit 1 & NRS1 &0.02854$\pm$0.00015 & 60149.25631$\pm$0.00012 &87.884$\pm$0.062 &26.49$\pm$0.62\\
         & &                 NRS2 &0.02860$\pm$0.00018 & 60149.25605$\pm$0.00013 &88.104$\pm$0.082 &28.79$\pm$0.96\\
         & Visit 2 &         NRS1 &0.02825$\pm$0.00012 & 60183.61414$\pm$0.00010 &88.134$\pm$0.057 &29.14$\pm$0.68\\
         & &                 NRS2 &0.02774$\pm$0.00015 & 60183.61406$\pm$0.00014 &88.232$\pm$0.102 &30.36$\pm$1.30\\
         & Joint &           NRS1 &0.02837$\pm$0.00009 & 60149.25624$\pm$0.00011, 60183.61414$\pm$0.00011\tablenotemark{a}&88.036$\pm$0.040&28.05$\pm$0.44\\
         &  &                NRS2 &0.02807$\pm$0.00011 & 60149.25611$\pm$0.00013, 60183.61408$\pm$0.00014\tablenotemark{a}&88.169$\pm$0.065&29.58$\pm$0.80\\
        \hline
        Tiberius & Visit 1 & NRS1 &$0.02856\pm0.00015$&$60149.25623\pm$0.00012 &$87.893\pm0.062$ & $26.59^{+0.64}_{-0.61}$ \\
        & & NRS2 & $0.02840\pm0.00019$&$60149.25630\pm$0.00015&$88.104^{+0.087}_{-0.083} $& $28.80^{+1.04}_{-0.93}$\\
        &Visit 2& NRS1 &$0.02824\pm0.00012$ &$60183.61415\pm$0.00010 & $88.160^{+0.058}_{-0.056}$ & $29.44^{+0.72}_{-0.67}$\\
        & & NRS2 &$0.02624\pm0.00023$\tablenotemark{b} &$60183.61380^{+0.00015}_{-0.00014}$ & $88.359^{+0.114}_{-0.113}$&$32.06^{+1.65}_{-1.52} $ \\
    \hline
    \end{tabular}
    \tablenotetext{a}{We do not phase the two visits owing to the presence of the TTVs, therefore we independently fit the $T_0$, shown for visit 1 and visit 2 respectively. }
        \tablenotetext{b}{There is an offset between the \texttt{Tiberius} transit depths for NRS2 visit 2 and the white light curve $R_p/R_*$ value likely owing to the difference in the systematic noise models utilized in the spectroscopic and white light curves. See Section~\ref{sec:tiberius} for more details.}
    \label{tab:fitted-values}
\end{table*}    

\section{System Parameters}
\label{sec:orbit}

\subsection{Orbital Period}
Given the earlier than expected transit in visit 1, we attempted to update the orbital parameters for HD 15337~c. The original ephemeris of both \cite{Dumusque2019} and \cite{Gandolfi2019} used only three transits of HD 15337~c from October and November 2018 in TESS Sectors 3 and 4, but two more transits became available in the interim from TESS Sector 30 in October 2020. We initially assumed the nearly missed transit was caused by a stale ephemeris, which could be updated by including the more recent TESS data. We use the \texttt{lightkurve} \citep{lightkurve} package to query the Mikulski
Archive for Space Telescopes and download the two-minute cadence photometry from TESS Sectors 3, 4, and 30 for HD 15337 using the TESS Science Process Operations Center pipeline. We first jointly fit all five TESS transits available in these sectors assuming the same $i$, $a/R_{*}$, $R_{p}/R_{*}$, period, and initial time of transit $T_0$. We find that our derived period and $T_0$ are in good agreement with those available on ExoFOP as of 2024 September 10, which are the same as those used in our initial transit scheduling to center the initial observation. This indicates that the inaccurate time of transit used to center the first observation was not a product of incorrect user provided values, but astrophysical in nature. Therefore, we strongly caution against using those derived period and $T_0$ parameters for reasons discussed below.

We then tried to update the linear period and ephemeris using the combination of the five TESS transits and four JWST white light curves (two each for NRS1 and NRS2). For these joint fits, we fix the parameters of the instrumental model for the JWST light curves to those from the individual fits to each white light curve to reduce dimensionality, allow different  $i$, $a/R_{*}$, and $R_{\rm p}/R_{*}$ values to increase flexibility, and assume a common period and transit time to yield the possible update in these parameters. However, propagating this newly derived time of transit and period to the individual fits for the JWST light curves creates a $\sim$90 second early and $\sim$90 second late transit time for the two observations (the two detectors for each visit are in good agreement). This small timing offset between JWST visits separated by $\sim$ 34 days led to the discovery of large TTVs in the combined TESS, CHEOPS, and JWST light curves across a nearly 5 year baseline (see Section \ref{sec:TTVs}). This indicates that a linear period is not an accurate description of the orbit of this planet.  We choose not to update the period of the planet, since any linear ephemeris will not be accurate due to the presence of TTVs. The true TTV amplitude and period are also not yet known, so any given estimate of a linear period or $T_0$ value, even an updated one, will be incorrect by an undetermined amount. There is a strong possibility that future transit scheduling based on a linear ephemeris far in time from the JWST observations will suffer from the same issues as the original JWST scheduling due to an incorrect linear ephemeris relative to the TESS data. We discuss the presence and implication of these TTVs in further detail in Section~\ref{sec:TTVs}.

\subsection{Transit Timing Variations of HD 15337~c}\label{sec:TTVs}
As we were unable to determine a linear period for HD 15337~c owing to the apparent presence of TTVs, we instead fit for individual transit times in the four JWST light curves from the two detectors across two JWST transit observations, the available TESS transits, as well as archival CHEOPS observations of both planets from \cite{Rosario2024}. We then attempt to fit a TTV model to determine if the TTVs could be explained by either the known smaller interior planet in the system or the presence of any undetected companions.

We perform a joint fit to the TESS and CHEOPS data without assuming a Keplerian orbital period, with the same model framework as described in \cite{GreklekMcKeon_2025}. We constructed our photometric model using the \texttt{exoplanet} package \citep{exoplanet:joss} for the transit light curve component. Our transit model includes the stellar radius $R_*$, planet-to-star radius ratio $R_p$/$R_*$, impact parameter $b$, and scaled semimajor axis $a/R_*$ shared across TESS and CHEOPS data for HD 15337~b and c, respectively, along with individual transit mid-center times $t_{0_i}$. We allowed the midtime of each transit to vary individually using the \texttt{TTVOrbit} module of \texttt{exoplanet}.

We simultaneously modeled systematics in the TESS data due to instrumental and stellar variability with a Matern-3/2 kernel GP from \texttt{celerite2} \citep{exoplanet:foremanmackey17,exoplanet:foremanmackey18}. This kernel is commonly utilized for its ability to model quasi-periodic stellar variability signals \citep[e.g.,][]{Gan2022,Cointepas2024}. Systematic model parameters included a mean offset $\mu$, GP hyperparameters $\sigma$ and $\rho$ corresponding to the amplitude and timescale of quasi-periodic oscillations, and an error scaling term added in quadrature to the flux errors.

We download the detrended CHEOPS transit light curves from \cite{Rosario2024} and refit the data jointly with TESS. We include a mean flux offset term for each CHEOPS visit and an error scaling term as used by \cite{Rosario2024}, but find that the flux offset terms are not significant in the fit (Table \ref{tab:TESS_TTV_fit}). We fixed the planetary eccentricities and $\omega$ values and the stellar quadratic limb darkening parameters in the TESS and CHEOPS bandpasses to those reported from the joint transit+RV fit of \cite{Rosario2024}.

\begin{table*}
  \centering
  \caption{Prior and Posterior Distributions for HD 15337 TESS Fit with Individual Transit Midtimes\label{tab:TESS_TTV_fit}}
  \begin{tabular}{lccc}
    \hline
    \hline
    Parameter & Prior & \multicolumn{2}{c}{Posterior} \\
    \hline
    \multicolumn{4}{c}{\emph{TESS systematics parameters}} \\
    \hline
    Mean flux offset, $\mu$ & $\mathcal{N}(0.0,1.0)$ & \multicolumn{2}{c}{$9.68 ^{+5.32}_{-5.27} \times10^{-6}$} \\
    Log jitter, $\ln{s^2}$ & $\mathcal{U}(-18,2)$ & \multicolumn{2}{c}{$-8.67\pm 0.02$} \\
    $\sigma$ (ppt) & $\mathcal{U}(10^{-4},1)$ & \multicolumn{2}{c}{$0.076^{+0.07}_{-0.06}$} \\
    $\rho$ (days) & $\mathcal{U}(10^{-2},10^{2})$ & \multicolumn{2}{c}{$0.55^{+0.12}_{-0.10}$} \\
    \multicolumn{4}{c}{\emph{CHEOPS systematics parameters}} \\
    \hline
    Mean flux offset (visit 1), $\mu_1$ & $\mathcal{N}(0.0,1.0)$ & \multicolumn{2}{c}{$-1.39\pm1.46 \times10^{-5}$} \\
    Mean flux offset (visit 2), $\mu_2$ & $\mathcal{N}(0.0,1.0)$ & \multicolumn{2}{c}{$1.76^{+10.61}_{-10.50} \times10^{-6}$} \\
    Mean flux offset (visit 3), $\mu_3$ & $\mathcal{N}(0.0,1.0)$ & \multicolumn{2}{c}{$4.78^{+12.51}_{-12.68} \times10^{-6}$} \\
    Mean flux offset (visit 4), $\mu_4$ & $\mathcal{N}(0.0,1.0)$ & \multicolumn{2}{c}{$-2.59^{+12.11}_{-12.03} \times10^{-6}$} \\
    Mean flux offset (visit 5), $\mu_5$ & $\mathcal{N}(0.0,1.0)$ & \multicolumn{2}{c}{$0.78^{+12.40}_{-12.20} \times10^{-6}$} \\
    Mean flux offset (visit 6), $\mu_6$ & $\mathcal{N}(0.0,1.0)$ & \multicolumn{2}{c}{$-1.76^{+14.14}_{-13.89} \times10^{-6}$} \\
    Log jitter, $\ln{s^2}$ & $\mathcal{U}(-18,2)$ & \multicolumn{2}{c}{$-8.48\pm 0.02$} \\
    \multicolumn{4}{c}{\emph{TESS + CHEOPS joint transit model parameters}} \\
    \hline
    && \emph{HD 15337~b} & \emph{HD 15337~c} \\
    $R_p/R_*$ & $\mathcal{U}(0.0,0.2)$ & $0.01887^{+0.00030}_{-0.00029}$ & $0.02708^{+0.00045}_{-0.00044}$ \\
    $a/R_*$ & $p(a/R_*\vert P,M_{*}, R_{*})$ & $13.50^{+0.23}_{-0.45}$ & $31.76^{+1.22}_{-1.19}$ \\
    Impact parameter, $b$ & $\mathcal{U}(0,1+R_p/R_*)$ & $0.198^{+0.124}_{-0.127}$ & $0.851\pm0.013$ \\
    \multicolumn{4}{c}{\emph{Measured transit times (BJD - 2,457,000)}} \\
    \hline
    && \emph{HD 15337~b} & \emph{HD 15337~c} \\
    $t_{0_1}$ & $\mathcal{U}$($t_{0_i}$ - 0.083,$t_{0_i}$ + 0.083) (days)$^\textsc{1}$ & $1387.6825 \pm 0.0053^\textsc{\textdagger}$ & $1397.3707\pm 0.0030^\textsc{\textdagger}$ \\
    $t_{0_2}$ & '' & $1392.4338 \pm 0.0028^\textsc{\textdagger}$ & $1414.5495 \pm 0.0018^\textsc{\textdagger}$ \\
    $t_{0_3}$ & '' & $1397.1921 \pm 0.0031^\textsc{\textdagger}$ & $1431.7292 \pm 0.0018^\textsc{\textdagger}$ \\
    $t_{0_4}$ & '' & $1401.9593 \pm 0.0056^\textsc{\textdagger}$ & $2118.9567 \pm 0.0032^\textsc{\textdagger}$ \\
    $t_{0_5}$ & '' & $1411.4585 \pm 0.0039^\textsc{\textdagger}$ & $2136.1403 \pm 0.0030^\textsc{\textdagger}$ \\
    $t_{0_6}$ & '' & $1416.2191 \pm 0.0030^\textsc{\textdagger}$ & $2462.5652 \pm 0.0000^{\ddagger}$  \\
    $t_{0_7}$ & '' & $1425.7296 \pm 0.0036^\textsc{\textdagger}$ & $2514.1058 \pm 0.0009^{\ddagger}$ \\
    $t_{0_8}$ & '' & $1430.4897 \pm 0.0033^\textsc{\textdagger}$ & $2531.2863 \pm 0.0012^{\ddagger}$ \\
    $t_{0_9}$ & '' & $1435.2356 \pm 0.0060^\textsc{\textdagger}$ & $3149.7560 \pm 0.0002^\textsc{*}$ \\
    $t_{0_{10}}$ & '' & $2120.1014 \pm 0.0044^\textsc{\textdagger}$ & $3184.1140 \pm 0.0002^\textsc{*}$ \\
    $t_{0_{11}}$ & '' & $2124.8566 \pm 0.0029^\textsc{\textdagger}$ & - \\
    $t_{0_{12}}$ & '' & $2134.3692 \pm 0.0043^\textsc{\textdagger}$ & - \\
    $t_{0_{13}}$ & '' & $2139.1284 \pm 0.0021^\textsc{\textdagger}$ & - \\
    $t_{0_{14}}$ & '' & $2462.5343 \pm 0.0009^{\ddagger}$ & - \\
    $t_{0_{15}}$ & '' & $2510.0968 \pm 0.0014^{\ddagger}$ & - \\
    $t_{0_{16}}$ & '' & $2524.3609 \pm 0.0005^{\ddagger}$ & - \\
    $t_{0_{17}}$ & '' & $2529.1210 \pm 0.0017^{\ddagger}$ & - \\
    \hline
    \multicolumn{4}{l}{\footnotesize{$^\textsc{1}$ The prior on each transit midtime is $\pm$ 2 hours from the time predicted by the linear ephemeris described in  Section~\ref{sec:orbit}}}\\
    \multicolumn{4}{l}{\footnotesize{$^\textsc{\textdagger}$ denotes a TESS transit, $^{\ddagger}$ denotes a CHEOPS transit, and $^{*}$ denotes a JWST transit.}}\\
\end{tabular}
\end{table*}

We list the priors and posteriors for our transit and systematics models in Table \ref{tab:TESS_TTV_fit}. We used wide uniform priors for most parameters but placed a Gaussian prior on $R_*$ and $a/R_*$ in our TESS and CHEOPS fits based on the stellar mass, radius, and planetary orbital periods of \cite{Rosario2024}. We used the \texttt{PyMC3} package \citep{exoplanet:pymc3} to sample the posterior distribution of our model with No U-Turn Sampling (NUTS), with four parallel chains run with 2500 burn-in steps and 2000 posterior sample draws. We confirmed that the chains evolved until the Gelman-Rubin statistic values are $<$ 1.01 for all parameters, and that the posterior distribution for each individual transit time is approximately unimodal. 

\begin{figure*}
\begin{centering}
\includegraphics[width=0.9\textwidth]{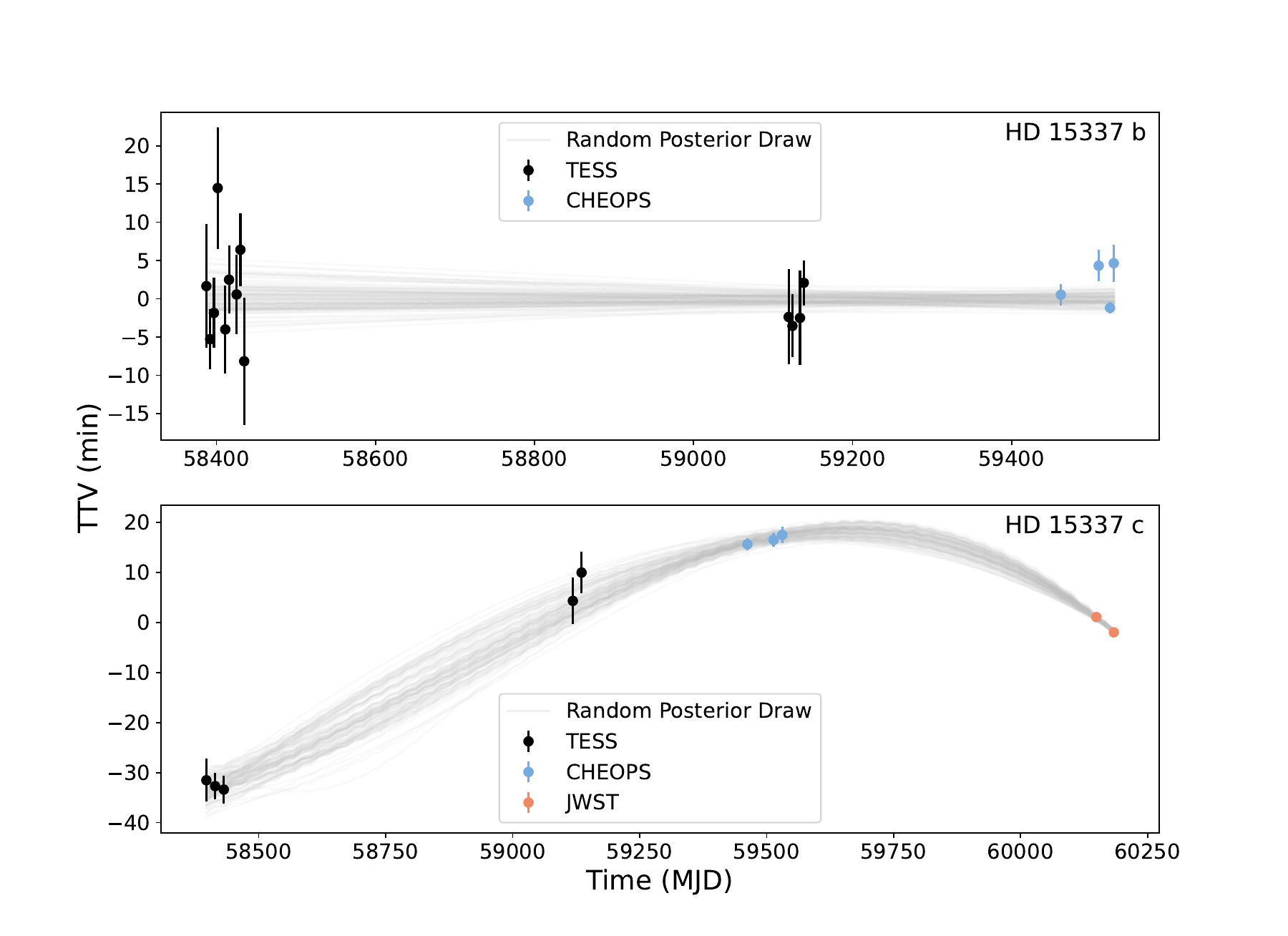} 
\caption{Observed TESS (black points) and CHEOPS (blue points) TTVs for HD 15337~b (top panel) and c (bottom panel), and JWST TTVs for HD 15337~c (red points), with 100 random draws from only one mode of the multimodal posterior distribution for a third planet exterior to HD 15337~c (gray lines). We caution that since only HD 15337~c currently exhibits TTVs and that there are only 10 observations to fit 10 model parameters, any solution is inherently poorly constrained and the true configuration of the system cannot be determined yet.}
\label{fig:TOI_402_TTVs}   
\end{centering}
\end{figure*}

In systems exhibiting TTVs, it is possible for best-fit transit shape parameters to shift when switching from a  model with a constant period to one that allows for independent/nonperiodic midtransit times \cite[e.g.,][]{GreklekMcKeon_2025}, but we found that our individual midtime fit produced consistent results to the constant period TESS model, which covers the longest baseline. We combined our TESS and CHEOPS midtimes with the error-weighted average JWST times from each detector for the two JWST visits to create our full TTV dataset, shown in Figure \ref{fig:TOI_402_TTVs}. We find no evidence of TTVs for the inner planet HD 15337~b, but with the new extreme-precision JWST transits anchoring the TTV data, we find significant peak-to-peak TTVs of ~30 minutes for HD 15337~c. It is likely that previous studies of the HD 15337 system \citep{Gandolfi2019,Dumusque2019,Rosario2024} did not identify TTVs because not enough of the TTV super-period \citep{Lithwick2012} had been sampled, and the TTV slope had not yet changed from positive to negative, allowing a linear ephemeris to adequately describe the observed transit times before the epochs of JWST observation. HD 15337~b has also been observed with JWST through the COMPASS program (Gagnebin et al. in prep), but even when including these data we do not see evidence of TTVs for HD 15337~b, so we reserve the detailed analysis of these data for future study.

We first attempted to determine if the two known transiting planets could produce the observed TTVs. HD 15337~b and c have a period ratio of 3.61, nowhere near any strong first- or second-order orbital resonance, which is typically responsible for causing TTVs \cite[e.g.,][]{Holczer_2016}. We computed the expected TTV amplitudes for the two known transiting planets using the \texttt{TTVFaster} package \citep{TTVFaster}, which uses analytic formulae derived from n-body simulations to model transit times that are accurate to first order in planet-star mass ratios and orbital eccentricities. We create 10,000 synthetic system configurations with planet mass, orbital period, and $T_0$ values drawn from the \cite{Rosario2024} distributions, initial eccentricities uniformly sampled from 0.0 to 0.5, and longitudes of periastron ($\omega$) randomly drawn from 0 to 2$\pi$. We consider up to 20th order resonances and simulate the system for 1000 days in each configuration with a time step of 0.2 days (less than 1/20 the period of HD 15337~b), and compute the maximum TTV amplitude from the \texttt{TTVFaster} model output for each configuration. We find that with the 2 planet model, the median TTV amplitude for HD 15337~c is 6 seconds, with a maximum of 15 seconds. This is incompatible with the TTV observations, and indicative of an additional body in the system perturbing the orbit of HD 15337~c. 

Given that our JWST observations revealed strong TTVs for HD 15337~c (see Figure \ref{fig:TOI_402_TTVs}) but there is no current evidence for TTVs in HD 15337~b, we perform a preliminary retrieval on the TTVs using a three-planet model that assumes the third planet is exterior to HD 15337~c. If there is a previously undetected planet between HD 15337~b and c, it would be more likely to also induce TTVs on HD 15337~b due to its proximity, and would more likely have been discovered by the RV analyses of \cite{Dumusque2019}, \cite{Gandolfi2019}, or \cite{Rosario2024}. A low-mass planet on an exterior orbit to HD 15337~c may have plausibly evaded detection in the RV observations, and a planet at the 2:1 resonance with HD 15337~c or beyond would not be transiting if it is coplanar with HD 15337~c. This is because the impact parameter of HD 15337~c is 0.851 (see Table \ref{tab:TESS_TTV_fit}), so a coplanar exterior planet with an orbital period of $\sim$ 34 days or more would have an impact parameter of at least 1.36. 

For these reasons we perform a preliminary retrieval on the TTV data with this exterior three-planet model, but caution that a full TTV model fit is incomplete with the current data and that our constraints do not represent the bona fide detection of a planet candidate. This is because the TTV super-period is not yet fully sampled, and any model that produces TTVs for HD 15337~c but not HD 15337~b requires 10 free parameters to describe the 10 observed transit times of HD 15337~c and is therefore poorly constrained. 

We construct and sample the three-planet TTV model using the same framework as \cite{GreklekMcKeon_2025}. We build the model using the \texttt{TTVFast} \citep{TTVFast} package, an $n$-body code that uses a symplectic integrator with a Keplerian interpolator to calculate transit times in multi-planet systems. We use Gaussian priors on the planet masses from \cite{Rosario2024}, uniform priors on the orbital period and $T_0$ parameters $\pm$ 1 day from the values reported in \cite{Rosario2024}, and re-parameterize eccentricity and $\omega$ as $\sqrt{e}cos{\omega}$ and $\sqrt{e}sin{\omega}$ with uniform priors from -1 to 1. For the exterior third planet, we allow orbital periods from 17.5 to 55 days, covering beyond the 3:1 resonance with HD 15337~c. In addition, we use a uniform prior in planet mass from 0 to 30 Earth masses, and the same e/$\omega$ prior and parameterization as for HD 15337~b and c. We initialize 1500 walkers for the 15 total model parameters centered at the \cite{Rosario2024} values for HD 15337~b and c while spreading the initial walker positions for the third planet uniformly across the parameter distribution, and use the \texttt{emcee} package to sample for 10$^5$ steps before sorting all the walkers by log probability, taking the highest likelihood 10\% of walkers, and then sampling for 10$^6$ steps. 

The purpose of this fit is to provide an initial exploration of the parameter space and determine the highest likelihood peaks of the posterior space without attempting to achieve convergence, since we found that even when allowing the MCMC to fully explore the parameter space for $> 10^6$ steps, the fit does not converge and the posterior is multi-modal. The posterior distribution of the orbital period for the external third planet has peaks near several orbital resonance locations (e.g., 3:2, 2:1, 3:1) relative to HD 15337~c. This is not unexpected, since the total number of TTV data points for HD 15337~c is equal to the number of free parameters used to model the TTVs, and the model is therefore poorly constrained. We cannot yet confidently determine which of these near-resonant states for the third planet is most likely. Given the limitations of the TTV model based on current data, we chose to briefly explore the current modes of the TTV posterior distribution, show one plausible example case, and reserve deeper analysis for future work. Since our unconverged posterior results in poorly estimated parameter uncertainties, we do not report the full posterior distribution from this procedure.

We then perform a representative example TTV fit with the framework described above, but instead we initialize the walkers for the period of the third planet near the 2:1 resonance with HD 15337~c at a period of 34.4 days and run the MCMC sampler for $10^5$ steps. We emphasize that this is an arbitrary choice, and a third planet near other exterior resonances with HD 15337~c such as the 3:1 and 3:2 provides a similar consistency with the data, but we cannot yet favor one orbital location over the others. In the 2:1 case, the posterior distribution for the third planet has an orbital period of 34.42$^{+0.07}_{-0.05}$ days with a planet mass of 2.8$^{+2.0}_{-1.4}$ $M_{\oplus}$, which exhibits good agreement with the observed TTVs (Figure \ref{fig:TOI_402_TTVs}), though we caution that this is not the only possible solution based on current data, and that future work with more data is needed to improve the TTV fit.

If this planet is near the exterior 2:1 location, it could plausibly have evaded previous detection in both transit and RV observations. A 2.8 $M_{\oplus}$ coplanar planet near the 2:1 resonance with HD 15337~c would have an impact parameter of 1.36 and an RV semi-amplitude of 0.6 m/s, smaller than the jitter for all RV datasets reported in \cite{Rosario2024}. This would represent a relatively rare system architecture in which the inner two planets of similar masses are accompanied by a much lower mass exterior companion, against the traditional peas-in-a-pod framework of compact multi-planet systems \citep{Millholland_2017, Weiss_2018}. We emphasize that this is only an example of one possible solution. At least one additional transit observation is required to obtain more observations than free parameters for the external third planet case, which could reveal the true location of the third body causing the TTVs in the HD 15337 system, though several more transits will likely be needed to produce strong mass and orbital period constraints. Thankfully, HD 15337 was reobserved by TESS in October 2025 for sector 97, and will be reobserved again in June through August 2026 for sectors 105, 106, and 107. These observations, combined with ground-based follow-up to fill in the long gaps between TESS observations, could obtain a more complete sampling of the TTV curve. Future work with this additional data, combined with potential improvements in transit timing precision from jointly fitting all available TESS, CHEOPS, JWST, and ground-based transit data, could lead to better constraints on the true architecture of the HD 15337 system. 

Recently, the low-mass (0.262 $^{+0.019}_{-0.018} M_{\odot}$) companion to HD 15337 \citep{Rosario2024} was further characterized via astrometry \citep{An2025}, with an orbital period of 760 $^{+550}_{-280}$ days and separation of 88 $^{+38}_{-23}$ au. Given that we have not yet observed more than one full TTV super-period for HD 15337~c over our observational baseline of $\sim$5 years, it is possible that some small portion of the observed TTVs are caused by the light travel time effect (LTT effect, also known as the R$\o$mer effect) due to the motion of HD 15337 around its barycenter relative to the distant, low-mass companion, although the precise magnitude of this effect cannot be accurately determined given the large uncertainty on the orbital period of the companion. However, the LTT effect cannot be responsible for the large-amplitude long-period TTVs of HD 15337~c, because any TTV caused by the LTT effect must also cause similar long-period TTVs for HD 15337~b, which we do not observe (Figure \ref{fig:TOI_402_TTVs}). Therefore the TTVs of HD 15337~c are most likely caused by a closer, non-transiting external companion.

\subsection{Eccentricity}
Given that \cite{Rosario2024} detected the eccentricity of HD 15337~c at $<$3$\sigma$, we also tested the effect of fixing the eccentricity to zero on our best-fit parameters and transmission spectra for both the \texttt{Eureka!}\ and \texttt{Tiberius} reductions. While the time of transit, $R_{\rm p}$/$R_{*}$, and transmission spectra in all cases were well within 1$\sigma$ of the values regardless of whether the eccentricity and $\omega$ were fixed to the values in Table~\ref{table:system}, fixed to circular, or allowed to vary, the $a$/R$_{*}$ and inclination did differ by $>$3$\sigma$. Assuming a circular orbit, we get $a/R_*$  values for visit 1 of 28.75$\pm$0.67 and 31.21$\pm$1.06 and for visit 2 of 31.61$\pm$0.74 and 32.98$\pm$1.48 for NRS1 and NRS2, respectively (joint values of 30.43$\pm$0.48 for NRS1 and 32.09$\pm$0.85 for NRS2) and inclinations for visit 1 of 88.212$\pm$0.052$^\circ$ and 88.395$\pm$0.070$^\circ$ and for visit 2 of  88.422$\pm$0.049$^\circ$ and 88.508$\pm$0.090$^\circ$ for NRS1 and NRS2, respectively (joint values of 88.340$\pm$0.034$^\circ$ for NRS1 and 88.452$\pm$0.055$^\circ$ for NRS2) for the \texttt{Eureka!}\ reduction. The \texttt{Tiberius} fits to the individual visits agree to better than 1$\sigma$ with those from \texttt{Eureka!}. Therefore, while the best fit astrophysical values differ depending on the assumptions of a circular or eccentric orbit, the two reductions agree as to the differences, and the derived spectra are not dependent on the eccentricity of the orbit.

\section{Transmission Spectrum of HD 15337~\MakeLowercase{c}} 
\label{sec:spectrum}
\begin{figure*}
    \centering
    \includegraphics[width=\textwidth]{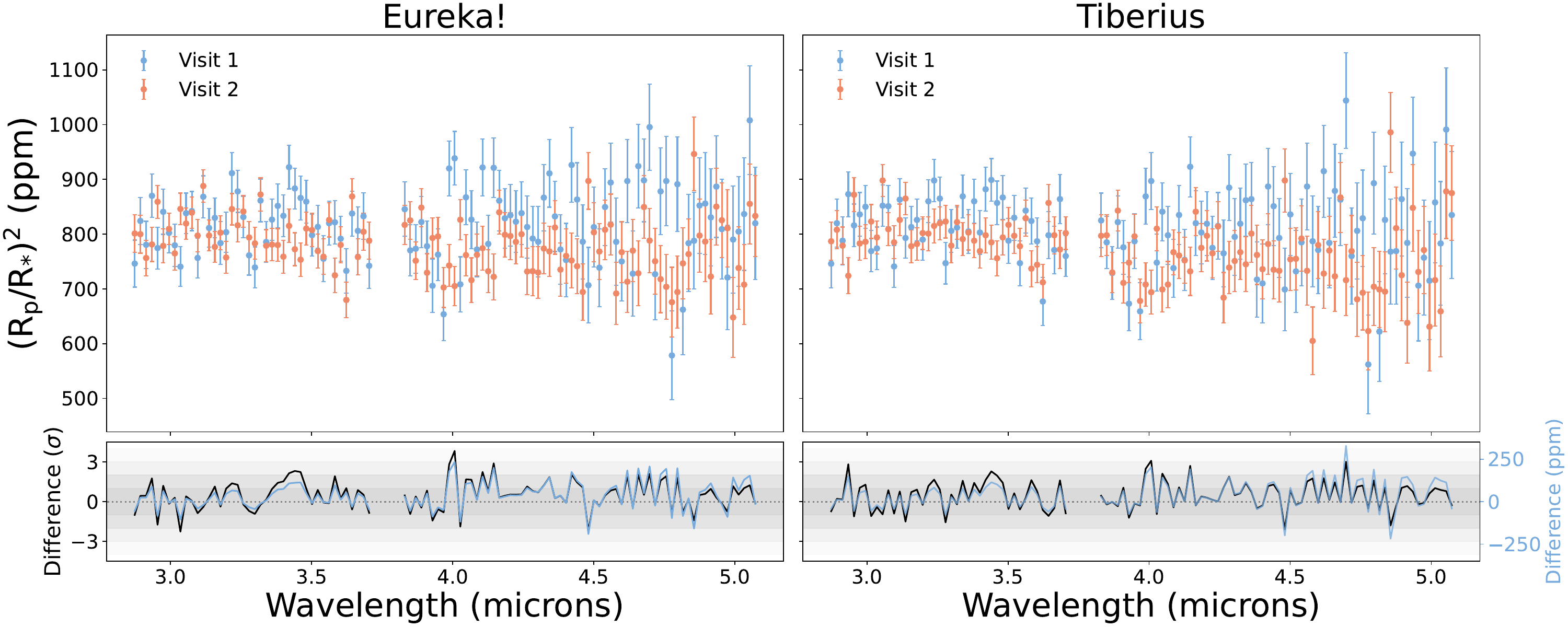}
    \caption{Spectra for both visits for each of the \texttt{Eureka!}\ (left) and \texttt{Tiberius} (right) reductions. We show the differences between the two visits within a reduction pipeline on the bottom panels in units of sigma (black) and ppm (blue). While there are some differences between the visits, the points generally agree, with median differences of $<$1$\sigma$. Only two points in the \texttt{Eureka!} spectra and  one point in the \texttt{Tiberius} spectra deviate by more than 3$\sigma$ between visits.}
    \label{fig:individual}
\end{figure*}

\begin{figure*}
    \centering
    \includegraphics[width=\textwidth]{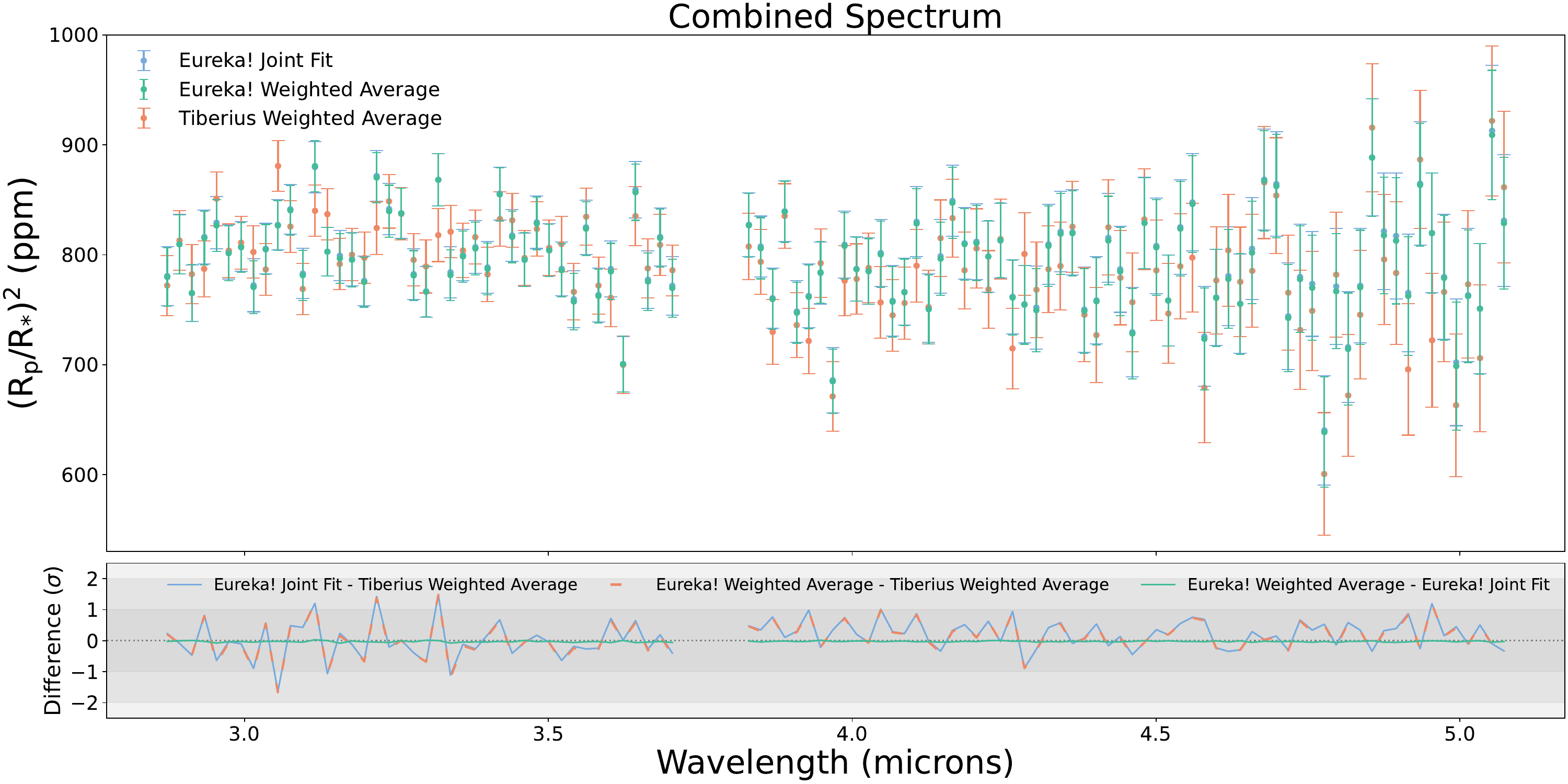}
    \caption{Combined spectra for the two visits of HD 15337~c. We show the error weighted average for the \texttt{Tiberius} reduction, and both the jointly fit and error weighted average for the \texttt{Eureka!}\ reduction. All three versions of the combined spectrum are in good agreement, with the median difference between the \texttt{Tiberius} reduction and \texttt{Eureka!} joint fit of $\sim$17 ppm and a maximum difference of $\sim$1.49$\sigma$.}
    \label{fig:joint}
\end{figure*}

\begin{figure*}
    \centering
    \includegraphics[width=.5\textwidth]{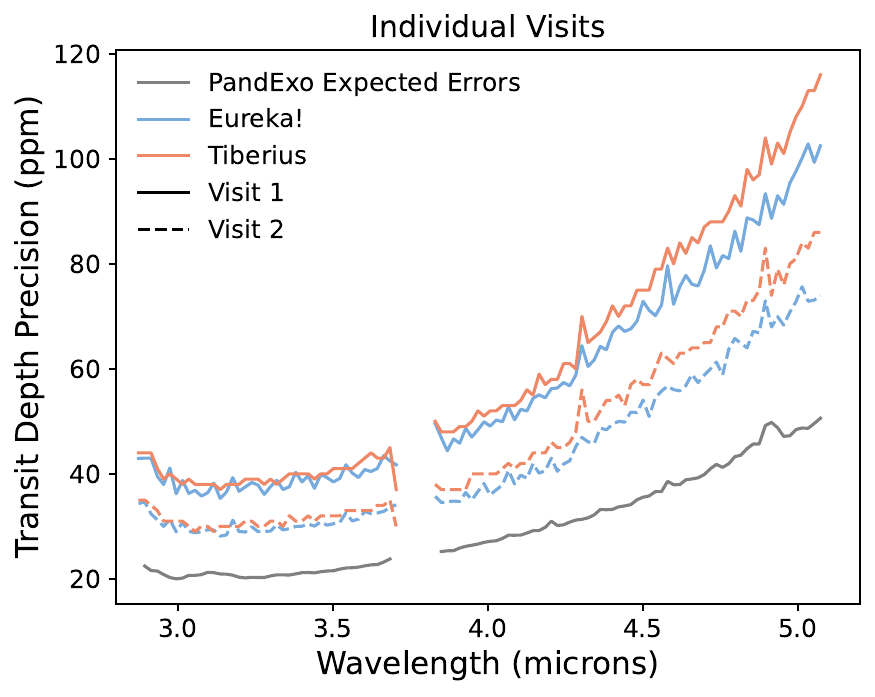}\includegraphics[width=.5\textwidth]{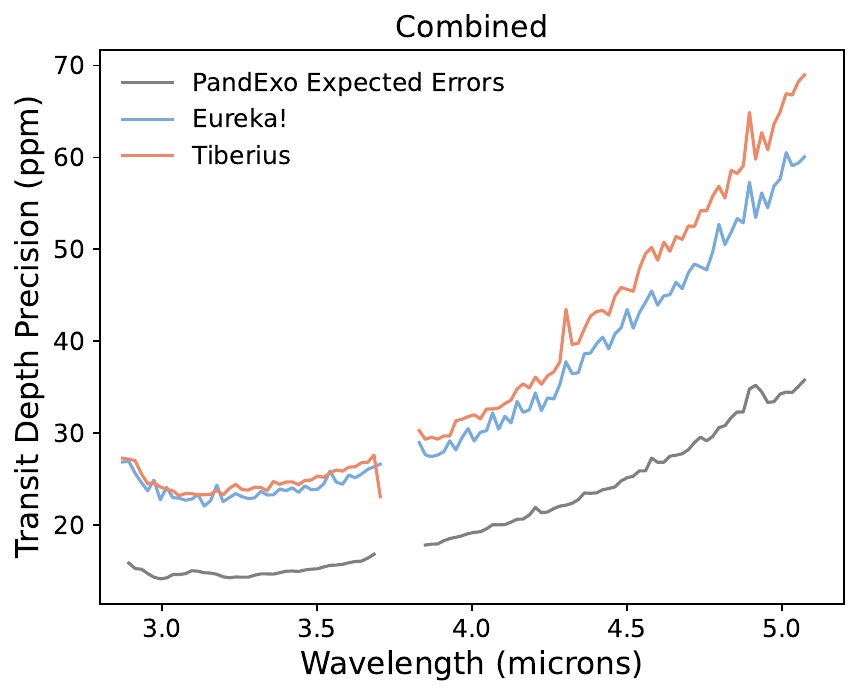}
    \caption{The expected errors on our transmission spectra from \texttt{PandExo} \citep{Batalha2017} with the measured errors for both reductions for the individual visits (left) and combined spectra (right). We achieve markedly better precisions for visit 2 compared to visit 1, likely owing to the lack of adequate pre-transit baseline for visit 1. We achieve median precisions for the \texttt{Eureka!}\ reductions for visit 1 of $\sim$39 ppm and $\sim$68 ppm and for visit 2 of $\sim$30 ppm and $\sim$51 ppm for NRS1 and NRS2 respectively. The precisions for the \texttt{Tiberius} reductions for visit 1 of $\sim$40 ppm and $\sim$72 ppm and for visit 2 of $\sim$31 ppm and $\sim$56 ppm for NRS1 and NRS2, respectively, are comparable to those of the \texttt{Eureka!}\ reduction. }
    \label{fig:PandExo}
\end{figure*}

We generated the transmission spectra of HD 15337~c using both of the aforementioned pipelines. We first fit each visit separately and present the spectra from these fits in Figure~\ref{fig:individual} and present the combined spectra in Figure~\ref{fig:joint}. There are a small number of points that deviate by more than 3$\sigma$ between the two visits within a reduction (two points for the \texttt{Eureka!}\ reduction and one point for the \texttt{Tiberius} reduction). However, the two reductions are in good agreement with a median difference between the reductions of 26 ppm for visit 1 and 20 ppm for visit 2. We compare our achieved per-point precision to that predicted by \texttt{PandExo} (Figure~\ref{fig:PandExo}) and find on average, in 30 pixel (R$\sim$200) bins for the \texttt{Eureka!}\ reductions, median precisions of $\sim$1.85$\times$ the expected \texttt{PandExo} precision for NRS1 and $\sim$1.99$\times$ the expected \texttt{PandExo} precision for NRS2 for visit 1 and $\sim$1.44$\times$ for NRS1 and $\sim$1.49$\times$ for NRS2 for visit 2. For the \texttt{Tiberius} reductions, we achieve median precisions of $\sim$1.88$\times$ the expected \texttt{PandExo} precision for NRS1 and $\sim$2.11$\times$ the expected \texttt{PandExo} precision for NRS2 for visit 1 and $\sim$1.47$\times$ for NRS1 and $\sim$1.66$\times$ for NRS2 for visit 2. The substantially lower precision in visit 1 highlights the likely importance of acquiring adequate pre-transit baseline in transmission spectrum precision. This may be especially crucial for NIRSpec observations with few numbers of groups where we may expect ramps to be present at the start of observations, which could be removed by trimming in the presence of adequate pre-transit baseline. We also tested the effect of fitting for the limb darkening coefficients using the \texttt{Eureka!} reduction of visit 2, as it is likely less affected by systematics than visit 1, and found consistent transmission spectra at a level better than 1.8$\sigma$ per point.

The two reductions are in good agreement for both the individual (Figure~\ref{fig:individual}) and combined spectra (Figure~\ref{fig:joint}), with median differences between the reductions of $\sim$26 ppm for visit 1, $\sim$20 ppm for visit 2 and $\sim$17 ppm for the combined spectra. Moreover, the spectra from the two different visits are in good agreement (Figure~\ref{fig:individual}), further justifying our combined spectrum. We also test the effect of jointly fitting our binned light curves by comparing the jointly fit transmission spectrum from \texttt{Eureka!}\ to the error weighted average of the two individual \texttt{Eureka!}\ spectra and find good agreement (Figure~\ref{fig:joint}). Therefore, we use the jointly fit \texttt{Eureka!}\ reduction in subsequent analyses. 

\section{Modeling the Atmosphere of HD 15337\MakeLowercase{c}}\label{sec:modeling}

The transmission spectra of HD 15337~c (Figures \ref{fig:individual} and \ref{fig:joint}) are largely flat and featureless across all reductions and visits. As in previous papers in the COMPASS program, we interpret the spectra using a suite of nonphysical and thermochemical equilibrium models to assess the significance of any potential spectral features and place constraints on the atmospheric metallicity/mean molecular weight and opaque pressure level, which may represent the top of a gray cloud deck. We apply these analyses to the spectra from the individual visits from the two reductions, as well as the jointly fit \texttt{Eureka!}\ spectrum and the weighted average \texttt{Tiberius} spectrum, which we will refer to as the ``combined'' spectra for simplicity. 

\subsection{Nonphysical Models}\label{sec:parametric}

We construct nonphysical models that can largely explain the overall shape of the spectrum without any reference to physical processes, and fit these models to the data in a Bayesian retrieval framework using \texttt{pymultinest} \citep{Buchner2014pymultinest}. Note that previous COMPASS papers used \texttt{ultranest} \citep{buchnerUltraNestRobustGeneral2021} for this effort, which is an all-Python version of \texttt{pymultinest}. We test 5 models: (1) a flat line with a single parameter: the average transit depth across both detectors; (2) a step function with two parameters: the average transit depth across all spectroscopic bins in NRS2 and the offset between the two detectors; (3) a sloped line with two parameters: the slope and the y-intercept;  (4) a step function plus a single Gaussian in NRS1, a model which has five parameters: the Gaussian amplitude, central wavelength, and standard deviation, and the average NRS1 and NRS2 transit depths; and (5) a step function plus a single Gaussian in NRS2, with the same parameters as the previous model. For the NRS1 Gaussian, the NRS2 portion of the model spectrum is set to be flat and equal to the NRS2 transit depth parameter, and vice versa for the NRS2 Gaussian. The Gaussian models are useful for evaluating the significance of any spectral features. These models are generated on the same wavelength grid as the data. 
The retrieval framework enables the calculation and comparison of the log of the Bayesian evidence, $lnZ$, across the different models, which allows us to evaluate which model is statistically preferred. We apply this retrieval framework to the individual spectra from all reductions and visits, and to the two combined spectra. We consider wide uniform or log-uniform priors for all parameters and 1000 live points for each retrieval. 

\begin{figure*}
\centering
\includegraphics[width=\textwidth]{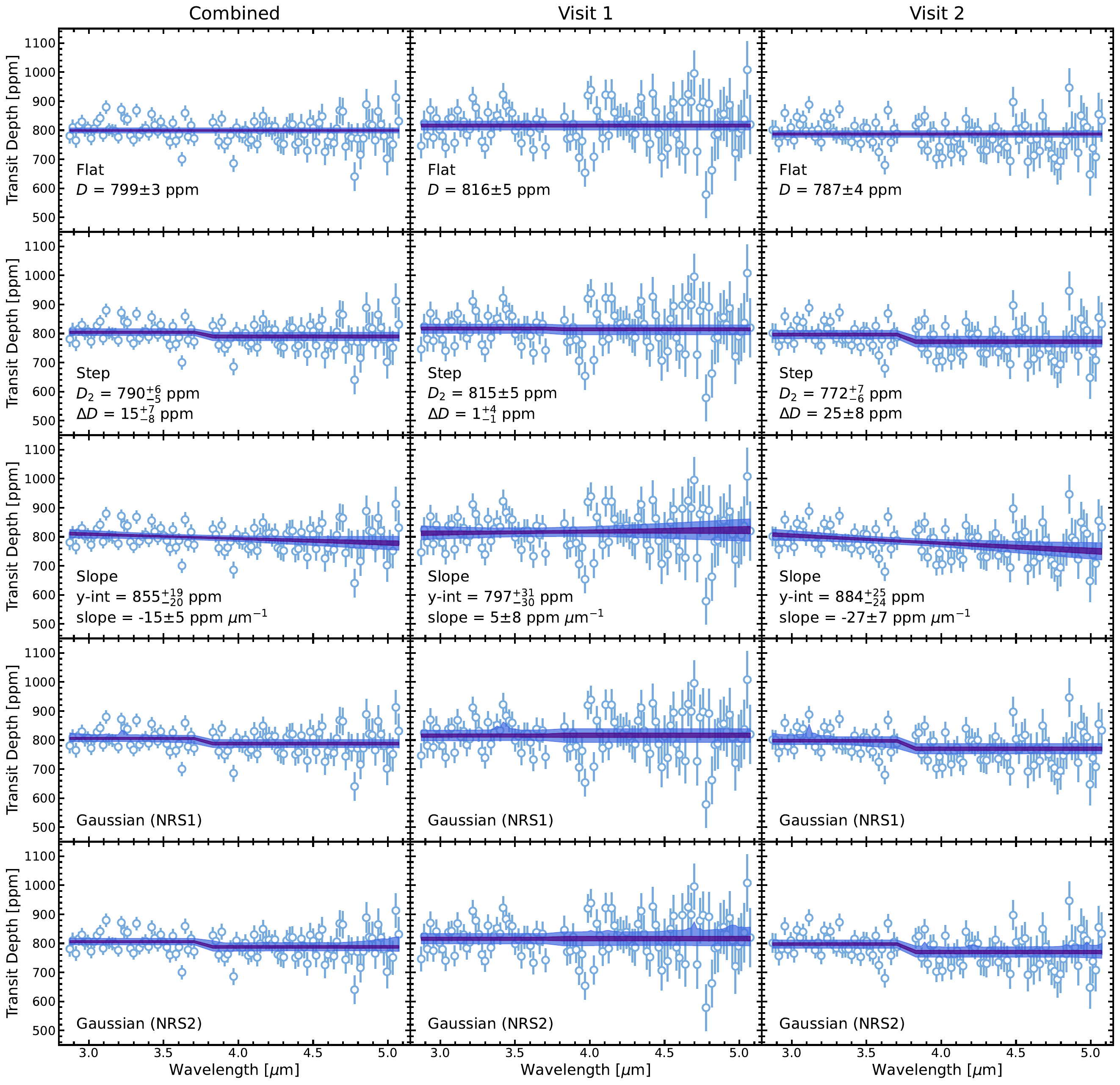}
\caption{Nonphysical model fits to the combined (left), visit 1 (middle), and visit 2 (right) \texttt{Eureka!}\ data reductions for the (top to bottom) flat, step function, slope, and NRS1 and NRS2 Gaussian models, respectively. For each panel, the data are given by the points, the 1$\sigma$ bounds of the best fit models are given by the inner, darker bands, and the 3$\sigma$ bounds are given by the outer, lighter bands. For the flat, step function, and slope models the best fit parameter values and associated uncertainties are given in the figure, where $D$ is the average transit depth across both detectors, $D_2$ is the average NRS2 transit depth, and $\Delta D$ is the offset between the NRS1 and NRS2 detectors; those of the Gaussian models are given in Table \ref{tab:gausstab}.}
\label{fig:unphysmod_eureka}
\end{figure*}

\begin{figure*}
\centering
\includegraphics[width=\textwidth]{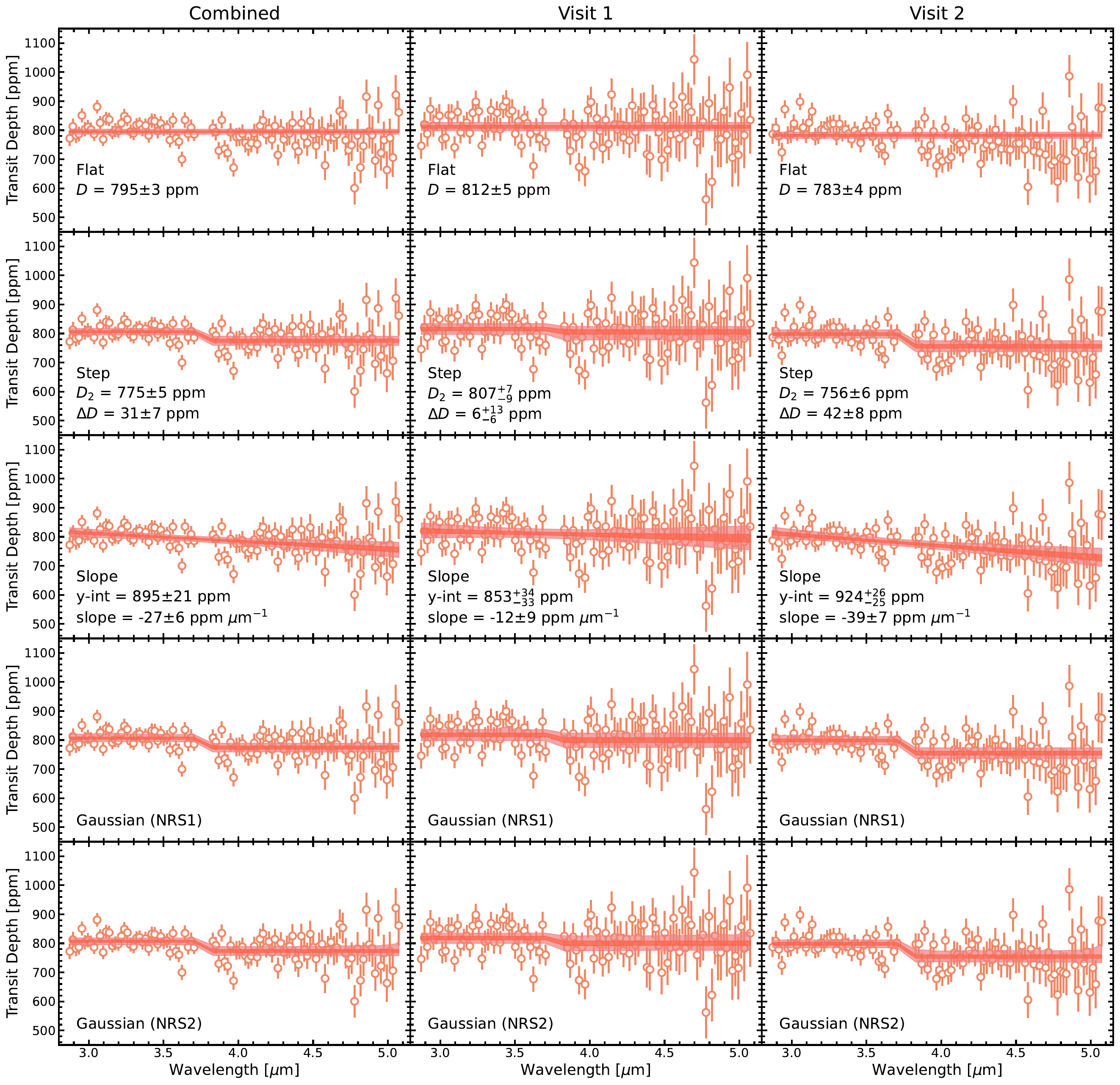}
\caption{Same as Fig. \ref{fig:unphysmod_eureka}, but for the \texttt{Tiberius} reductions. }
\label{fig:unphysmod_tiberius}
\end{figure*}

Our model fits and $lnZ$ values are shown in Figures \ref{fig:unphysmod_eureka} and \ref{fig:unphysmod_tiberius} and Table \ref{tab:lnz}, respectively. These reveal that the step function is significantly preferred ($\Delta lnZ>5$) over the flat line model for the combined and visit 2 \texttt{Tiberius} reductions and weakly preferred for the combined and visit 2 \texttt{Eureka!}\ reductions. The $lnZ$ values of the step function and the flat line model are similar for visit 1 of both reductions. While the step function is preferred over the slope model for all visits and reductions, the preference is weak across the board. The retrieved offsets in the step function and slopes in the slope model (see Figures \ref{fig:unphysmod_eureka} and \ref{fig:unphysmod_tiberius} for the values) are within 1-2$\sigma$ of each other across the different visits, respectively, with visit 2 data possessing steeper slopes and greater offsets between NRS1 and NRS2 compared to visit 1. The different behaviors between visits suggest that the offsets and/or slopes could be due to instrumental systematics, which has been seen in previous NIRSpec/G395H data sets (e.g., \citetalias{May2023}; \citealt{Wallack2024}). Finally, the Gaussian models, which are built on top of a step function, are not preferred over the step functions by themselves. In addition, the retrieved Gaussian parameters (Table \ref{tab:gausstab}) have large uncertainties. These findings indicate a lack of any significant spectral features in the data across all visits and reductions. In summary, our nonphysical modeling suggests that our 3-5 $\mu$m transmission spectrum of HD 15337~c is featureless and largely flat, with a possible slope or offset between NRS1 and NRS2 likely caused by instrumental systematics. 

\begin{table*}
\centering
\caption{$lnZ$ Values of the Nonphysical Models for the \texttt{Eureka!}\ (\texttt{E!}) and \texttt{Tiberius} (\texttt{T}) Combined (c), Visit 1 (v1), and Visit 2 (v2) Reductions}
\begin{tabular}{lcccccc}
\toprule
Model & \texttt{E!} (c) & \texttt{E!} (v1) & \texttt{E!} (v2) & \texttt{T} (c) & \texttt{T} (v1) & \texttt{T} (v2) \\
\hline
Flat line & -92 & -82 & -82 & -94 & -74 & -87 \\
Step function & -90 & -83 & -78 & -83 & -74 & -75 \\
Slope & -93 & -86 & -79 & -87 & -78 & -77 \\
Gaussian (NRS1) & -92 & -85 & -79 & -84 & -76 & -76 \\
Gaussian (NRS2) & -92 & -85 & -79 & -84 & -76 & -76 \\
\hline
\end{tabular}
\label{tab:lnz}
\end{table*}

\begin{table*}
\centering
\caption{Best-fit Parameters of the NRS1 and NRS2 Gaussian Models for the \texttt{Eureka!}\ (\texttt{E!}) and \texttt{Tiberius} (\texttt{T}) Combined (c), Visit 1 (v1), and Visit 2 (v2) Reductions}
\begin{tabular}{lcccccc}
\toprule
Parameter & \texttt{E!} (c) & \texttt{E!} (v1) & \texttt{E!} (v2) & \texttt{T} (c) & \texttt{T} (v1) & \texttt{T} (v2) \\
\hline
\multicolumn{7}{c}{Gaussian (NRS1)} \\
\hline
Amplitude (ppm) & 9$^{+35}_{-7}$ & 9$^{+35}_{-7}$ & 9$^{+33}_{-7}$ & 7$^{+29}_{-5}$ & 9$^{+32}_{-7}$ & 7$^{+30}_{-5}$ \\
Central wavelength ($\mu$m) & 4.92$^{+3.41}_{-3.39}$ & 5.24$^{+3.17}_{-3.52}$ & 4.83$^{+3.43}_{-3.26}$ & 5.17$^{+3.24}_{-3.55}$ & 5.02$^{+3.46}_{-3.52}$ & 4.84$^{+3.36}_{-3.33}$ \\
Standard deviation ($\mu$m) & 0.05$^{+1.29}_{-0.05}$ & 0.06$^{+1.37}_{-0.05}$ & 0.07$^{+1.37}_{-0.07}$ & 0.07$^{+1.5}_{-0.06}$ & 0.07$^{+1.73}_{-0.06}$ & 0.05$^{+1.37}_{-0.05}$ \\
NRS1 transit depth (ppm) & 805$\pm$4 & 815$^{+5}_{-6}$ & 797$\pm$5 & 806$\pm$4 & 817$^{+6}_{-7}$ & 798$\pm$5 \\
NRS2 transit depth (ppm) & 788$\pm$5 & 816$^{+8}_{-7}$ & 770$\pm$6 & 774$\pm$5 & 800$\pm$8 & 755$\pm$6 \\
\hline
\multicolumn{7}{c}{Gaussian (NRS2)} \\
\hline
Amplitude (ppm) & 8$^{+32}_{-6}$ & 8$^{+33}_{-6}$ & 8$^{+33}_{-6}$ & 8$^{+33}_{-6}$ & 9$^{+33}_{-7}$ & 8$^{+27}_{-6}$ \\
Central wavelength ($\mu$m) & 5.52$^{+2.92}_{-3.83}$ & 5.29$^{+3.14}_{-3.53}$ & 5.03$^{+3.29}_{-3.35}$ & 5.28$^{+3.12}_{-3.77}$ & 5.23$^{+3.12}_{-3.58}$ & 5.18$^{+3.05}_{-3.54}$ \\
Standard deviation ($\mu$m) & 0.09$^{+1.9}_{-0.08}$ & 0.06$^{+1.48}_{-0.06}$ & 0.07$^{+1.41}_{-0.06}$ & 0.06$^{+1.45}_{-0.06}$ & 0.07$^{+1.73}_{-0.07}$ & 0.07$^{+1.48}_{-0.07}$ \\
NRS1 transit depth (ppm) & 806$\pm$4 & 816$^{+6}_{-5}$ & 797$^{+5}_{-4}$ & 807$\pm$4 & 818$\pm$6 & 798$\pm$5 \\
NRS2 transit depth (ppm) & 787$\pm$5 & 815$\pm$8 & 770$^{+5}_{-6}$ & 773$^{+5}_{-6}$ & 799$^{+8}_{-9}$ & 754$^{+6}_{-7}$ \\
\hline
\end{tabular}
\label{tab:gausstab}
\end{table*}

\subsection{PLATON Retrievals}

Assuming thermochemical equilibrium for an atmosphere given a temperature and a set of elemental abundances can provide a first order limit on the atmospheric composition when the spectrum is largely featureless. Previous studies in the COMPASS program that encountered featureless spectra have used thermochemical modeling to place limits on the atmospheric metallicity/mean molecular weight and the opaque pressure level associated with an aerosol layer and/or the solid planetary surface. Here we conduct a similar exercise using a Bayesian retrieval framework, as in \citetalias{Teske2025}. 

We use the updated PLATON \citep[PLanetary Atmospheric Tool for Observer Noobs, version 6;][]{Zhang2025PLATON} retrieval code for this effort. PLATON 6 improves upon the previous PLATON versions \citep{Zhang2019PLATON,Zhang2020PLATON} in several ways, but for the purposes of this study the most important updates are (1) the addition of \texttt{pymultinest} as a sampling tool, (2) the increase in spectral resolution of the default opacities from 1000 to 20000, (3) the addition of up-to-date opacities for certain chemical species \citep[see Table 2 in][]{Zhang2025PLATON}, (4) the switch from using GGChem \citep{Woitke2018} to FastChem \citep{Stock2022} for calculating equilibrium chemical abundances, (5) the scaling of the sum of C and O rather than just O for a given metallicity, and (6) the use of solar abundances from \citet{Asplund2021} rather than \citet{Asplund2009}. We use the \texttt{pymultinest} sampler with the default R=20000 opacities and 1000 live points for each of our retrievals. 

Table \ref{tab:platonparams} gives the parameters we retrieve for and their priors. For the stellar radius $R_*$ and planet mass $M_p$ we use a Gaussian prior with the same central value and the larger of the 1$\sigma$ uncertainties (when they are asymmetric) as in Table \ref{table:system}. As PLATON fixes the retrieved planet radius $R_p$ to be that at 1 bar pressure in the atmosphere, and we are a priori uncertain about what pressures are being probed in transmission within 1-2 dex \citep[e.g.,][]{Grimm2018,Gao2020}, we use a wide uniform prior (2-3 R$_{\oplus}$) for $R_p$ to ensure we capture the correct radius at the relevant pressures. 

We further retrieve (1) the atmospheric metallicity with a log-uniform prior between 1 and 1000 $\times$ solar; (2) the opaque pressure level ($P_{\rm opaque}$) with a log-uniform prior between 10 $\mu$bar and 1 bar, representing an aerosol layer or the planet surface, with a limb fraction of 1 (e.g., no patchy clouds); and (3) an offset between NRS1 and NRS2 with a conservative uniform prior from -500 to +500 ppm. We assume a fixed isothermal temperature profile at the equilibrium temperature of HD 15337~c ($T_{eq}$ = 656 K) and an effective temperature for the star of 5131 K (Table \ref{table:system}). The C/O ratio of the atmosphere is assumed to be fixed at solar \citep[0.59;][]{Asplund2021}, as variations thereof have minimal impact on the atmospheric mean molecular weight at the temperature of HD 15337~c for the range of C/O values PLATON 6 allows (0.001 to 2). While PLATON 6 has been updated to include free chemical retrievals, we assume thermochemical equilibrium here since our goal is to rule out parameter space in atmospheric metallicity/mean molecular weight versus opaque pressure level, and also because we do not detect any statistically significant spectral features in our data (Section~\ref{sec:parametric}).

\begin{table}
    \centering
    \caption{Fitted Parameters and Priors in Our PLATON Retrievals}
    \noindent\begin{tabular}{lcc}
    \toprule
        Parameter & Prior Format & Prior Values\tablenotemark{a} \\
        \hline
         $R_{*}$ ($R_{\odot}$) & Gaussian & 0.008\\
         $M_p$ ($M_{\earth}$) & Gaussian & 1.302\\
         $R_p$ ($R_{\earth}$) & Uniform & [2,3]\\
         log[M/H ($\times$ Solar)] & Uniform  & [0,3] \\
         log[$P_{\rm opaque}$ (bars)] & Uniform &  [-5,0] \\
         Offset (ppm) & Uniform & [-500,500]\\
         \hline
    \end{tabular}
    \tablenotetext{a}{For Gaussian priors this gives the standard deviation. For uniform priors this gives the lower and upper bounds. }
    \label{tab:platonparams}
\end{table}

\begin{figure*}
\centering
\includegraphics[width=\textwidth]{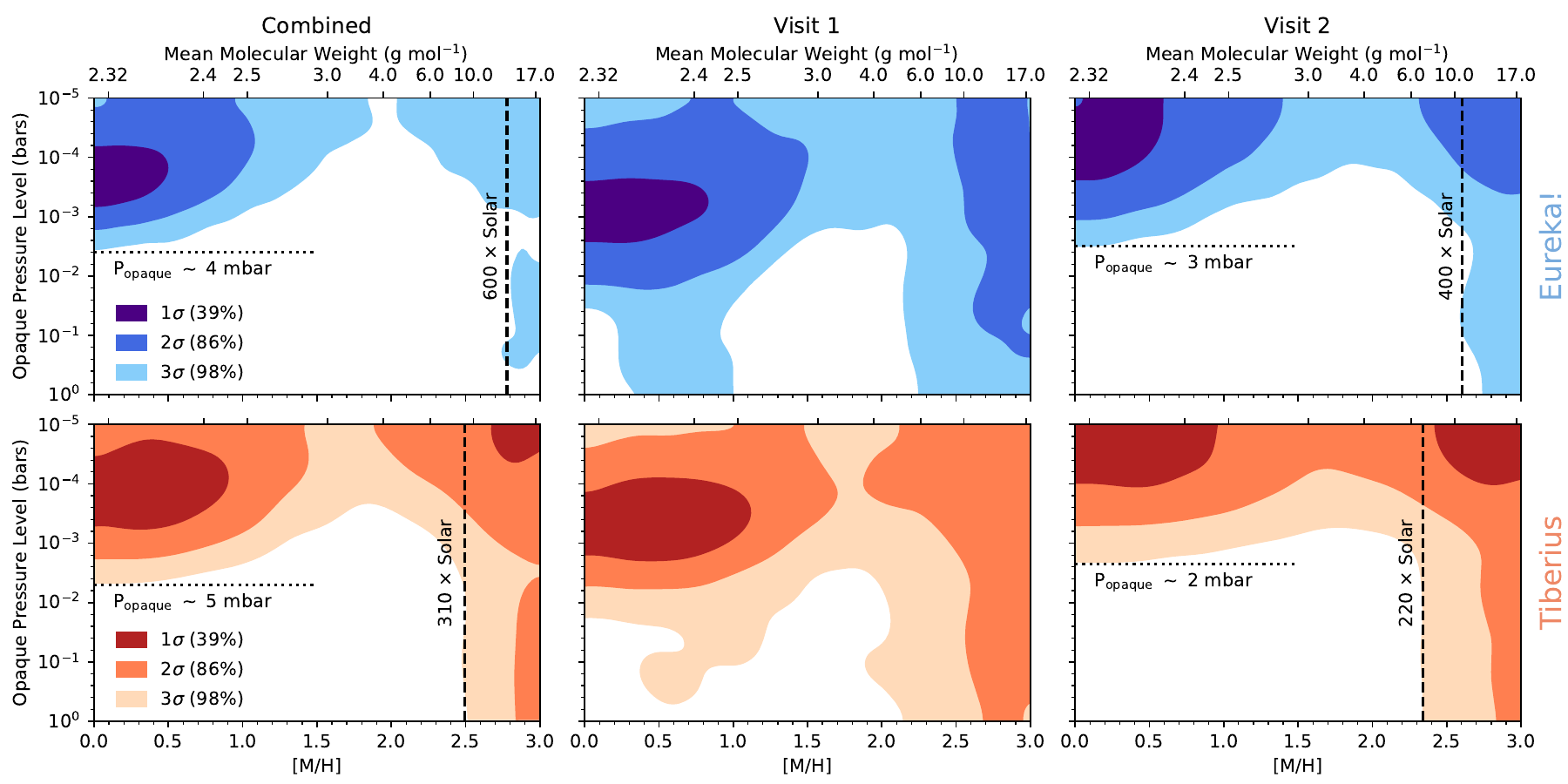}
\caption{$\sigma$-rejection contours in metallicity-opaque pressure level ($P_{\rm opaque}$) space from our PLATON retrievals for \texttt{Eureka!}\ (top) and \texttt{Tiberius} (bottom) reductions of the combined (left), visit 1 (middle), and visit 2 (right) data. The 1, 2, and 3$\sigma$ contours in 2D are indicated. The mean molecular weight at 1 bar in an isothermal atmosphere with HD 15337~c's equilibrium temperature is shown in the top axes and corresponds to the metallicities at the bottom axes, computed using PLATON assuming C/O = 0.59 and thermochemical equilibrium. Minimum allowed clear-atmosphere metallicities for the more constraining data sets are indicated by the vertical dashed lines, while the maximum allowed P$_{\rm opaque}$ values at lower metallicities are indicated by the horizontal dotted lines.}
\label{fig:metpcld}
\end{figure*}

\begin{figure*}
\centering
\includegraphics[width=\textwidth]{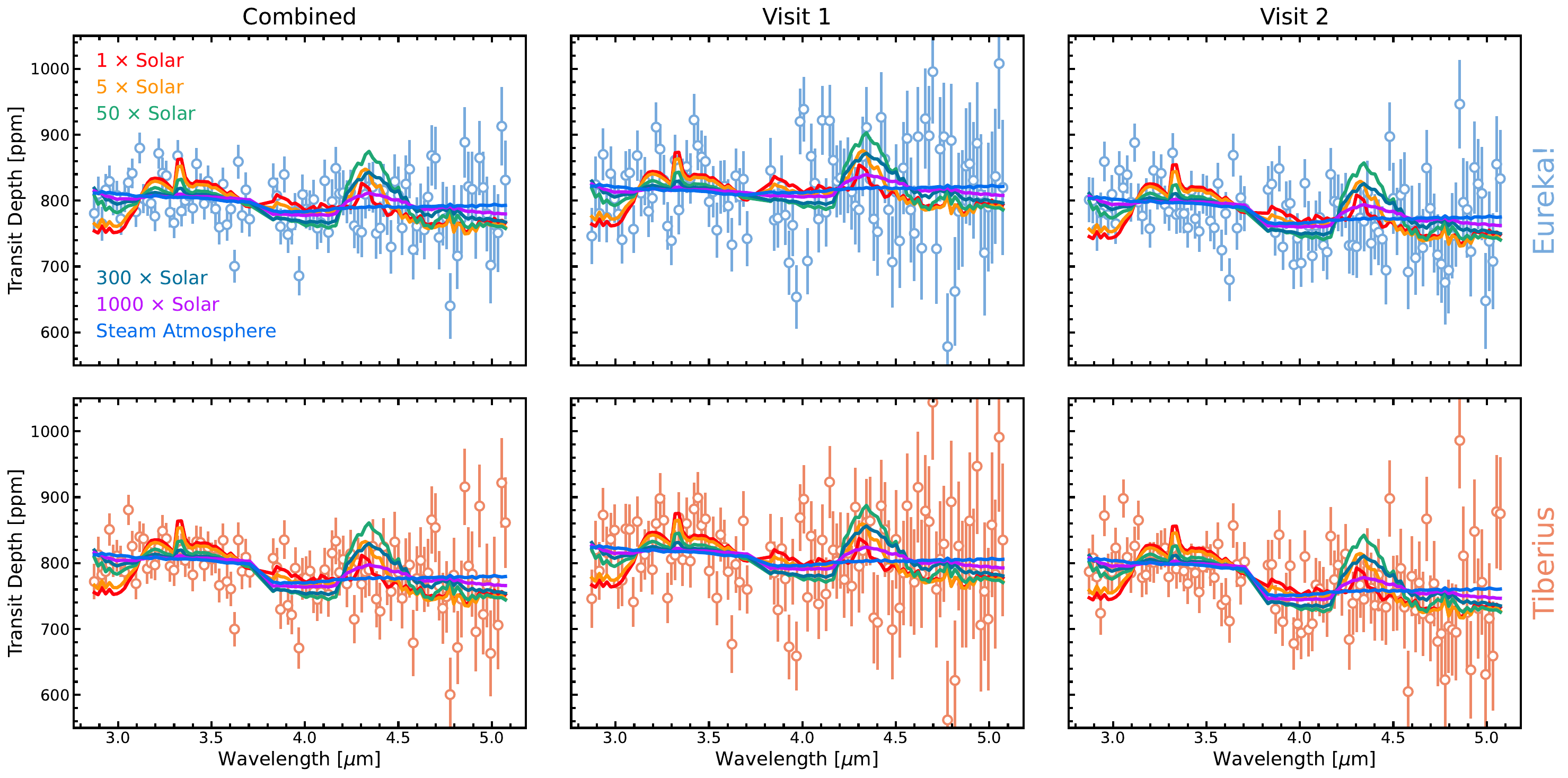}
\caption{Comparisons between representative thermochemical equilibrium atmospheric models (solar metallicity: red; 5 $\times$ solar: orange; 50 $\times$ solar: green; 300 $\times$ solar: dark blue; 1000 $\times$ solar: magenta; pure H$_2$O atmosphere: light blue) and the \texttt{Eureka!}\ (top) and \texttt{Tiberius} (bottom) reductions of the combined (left), visit 1 (middle), and visit 2 (right) data.} 
\label{fig:models}
\end{figure*}

We find that the data prefer high metallicities and/or high altitude aerosols, as shown in Figure \ref{fig:metpcld}. Specifically, we find that the combined spectra can rule out (at $>$3$\sigma$) metallicities $\leq$600 $\times$ solar (atmospheric mean molecular weight MMW $\leq13$ g mol$^{-1}$) for the \texttt{Eureka!}\ reduction and $\leq$310 $\times$ solar (MMW $\leq9$ g mol$^{-1}$) for the \texttt{Tiberius} reduction for opaque pressure levels greater than a few mbar. A greater range of metallicities is possible for lower opaque pressures. Most of the constraining power of the combined spectra arise from the higher precision visit 2 data, which by themselves can rule out $\leq$400 $\times$ solar and $\leq$220 $\times$ solar metallicity for the \texttt{Eureka!}\ and \texttt{Tiberius} reductions, respectively, for opaque pressure levels greater than a few mbar. In contrast, the much noisier visit 1 data have minimal constraining power.\footnote{The slightly tighter constraints on the opaque pressure level on the visit 2 data compared to the combined spectra could be caused by scatter introduced by the visit 1 data when combining the visits.} The relatively large difference in the metallicity constraints between largely consistent reductions have been seen previously \citep[e.g.][]{Alderson2024} and suggest that these inferences could be highly sensitive to even minor differences in transit depths and their precision as a function of wavelength (Figures \ref{fig:joint} and \ref{fig:PandExo}). 

\begin{table*}
\centering
\caption{Reduced Chi-square Values of the Thermochemical Equilibrium Model Fits for the \texttt{Eureka!}\ (\texttt{E!}) and \texttt{Tiberius} (\texttt{T}) Combined (c), Visit 1 (v1), and Visit 2 (v2) Reductions Shown in Figure \ref{fig:models}}
\begin{tabular}{lcccccc}
\toprule
Model & \texttt{E!} (c) & \texttt{E!} (v1) & \texttt{E!} (v2) & \texttt{T} (c) & \texttt{T} (v1) & \texttt{T} (v2) \\
\hline
1 $\times$ solar & 2.19 & 1.57 & 1.86 & 2.07 & 1.44 & 1.75 \\
5 $\times$ solar & 2.09 & 1.54 & 1.79 & 1.92 & 1.39 & 1.65 \\ 
50 $\times$ solar & 2.15 & 1.60 & 1.78 & 1.87 & 1.41 & 1.59 \\
300 $\times$ solar & 1.82 & 1.51 & 1.55 & 1.60 & 1.32 & 1.41 \\
1000 $\times$ solar & 1.62 & 1.46 & 1.39 & 1.45 & 1.28 & 1.31 \\
Steam atmosphere & 1.55 & 1.46 & 1.32 & 1.40 & 1.27 & 1.26 \\
\hline
\end{tabular}
\label{tab:redchisqr}
\end{table*}

Our retrieval results can be understood in light of how atmospheric spectra vary with increasing metallicity at HD 15337~c's equilibrium temperature ($T_{eq}$ = 656 K). Figure \ref{fig:models} shows several representative model spectra computed using PLATON assuming clear atmospheres, C/O = 0.59, and thermochemical equilibrium at HD 15337~c's $T_{eq}$ for a variety of metallicities, as well as a pure steam atmosphere. These are fitted to the data using \texttt{scipy.minimize} assuming the NRS2 transit depth and an offset between the NRS1 and NRS2 detectors as free parameters. We find that low metallicity ($<$100 $\times$ solar), clear atmospheres possess large CH$_4$ and CO$_2$ features at 3.3 and 4.3 $\mu$m, respectively, that stick out above the flatness of the spectra. This results in high $\chi^2/N$ values (Table \ref{tab:redchisqr}) and enable the retrievals to rule out these low metallicity cases. Meanwhile, higher metallicity atmospheres, including those dominated by non-H$_2$/He gases, offer better fits due to their smaller spectral features brought on by smaller scale heights. In other words, our data prefer atmospheric models that minimize spectral feature sizes, consistent with the findings from our nonphysical modeling in Section \ref{sec:parametric}. In addition, we see a more subtle effect at low metallicities where models with metallicities $\sim$5 $\times$ solar are sometimes more preferred compared to lower and (slightly) higher metallicity models (Figure \ref{fig:metpcld}). This is due to the 5 $\times$ solar models having a slightly smaller \ce{CH4} feature than lower metallicity models while simultaneously possessing a smaller \ce{CO2} feature than higher metallicity models (Figure \ref{fig:models}) and are therefore slightly flatter than either of those metallicity ranges. 

\section{The Landscape of Sub-Neptune Atmospheric Detections}\label{sec:discussion}

\begin{figure}
\begin{centering}
\includegraphics[width=0.5\textwidth]{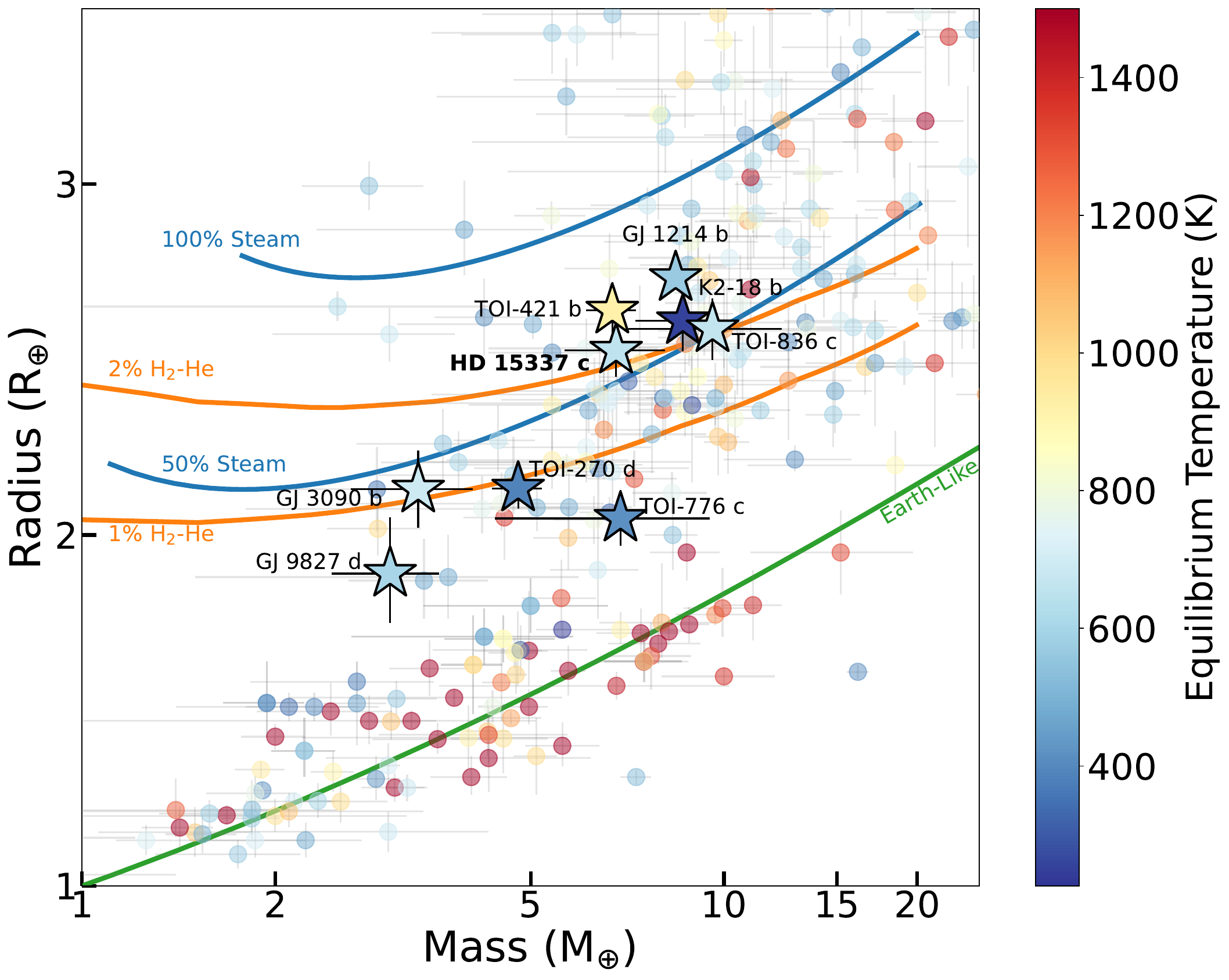} 
\caption{Mass-radius diagram of planets between 1 and 3.5 R$_{\oplus}$ colored by equilibrium temperature. We show planets with published JWST spectra and  R$_{\rm p}$ $>$ 1.7 R$_{\oplus}$ as stars with background points from the NASA Exoplanet Archive with measured masses and radii. We show curves of constant composition for an Earth-like planet from \cite{Zeng2016}, for steam from \cite{Aguichine2021} assuming a 600 K planet, and for H$_{2}$-He from \cite{Fortney2013} assuming a 600 K planet at an age of 5 Gyr.}
\label{fig:mr}   
\end{centering}
\end{figure}

\begin{figure*}
\begin{centering}
\includegraphics[width=\textwidth]{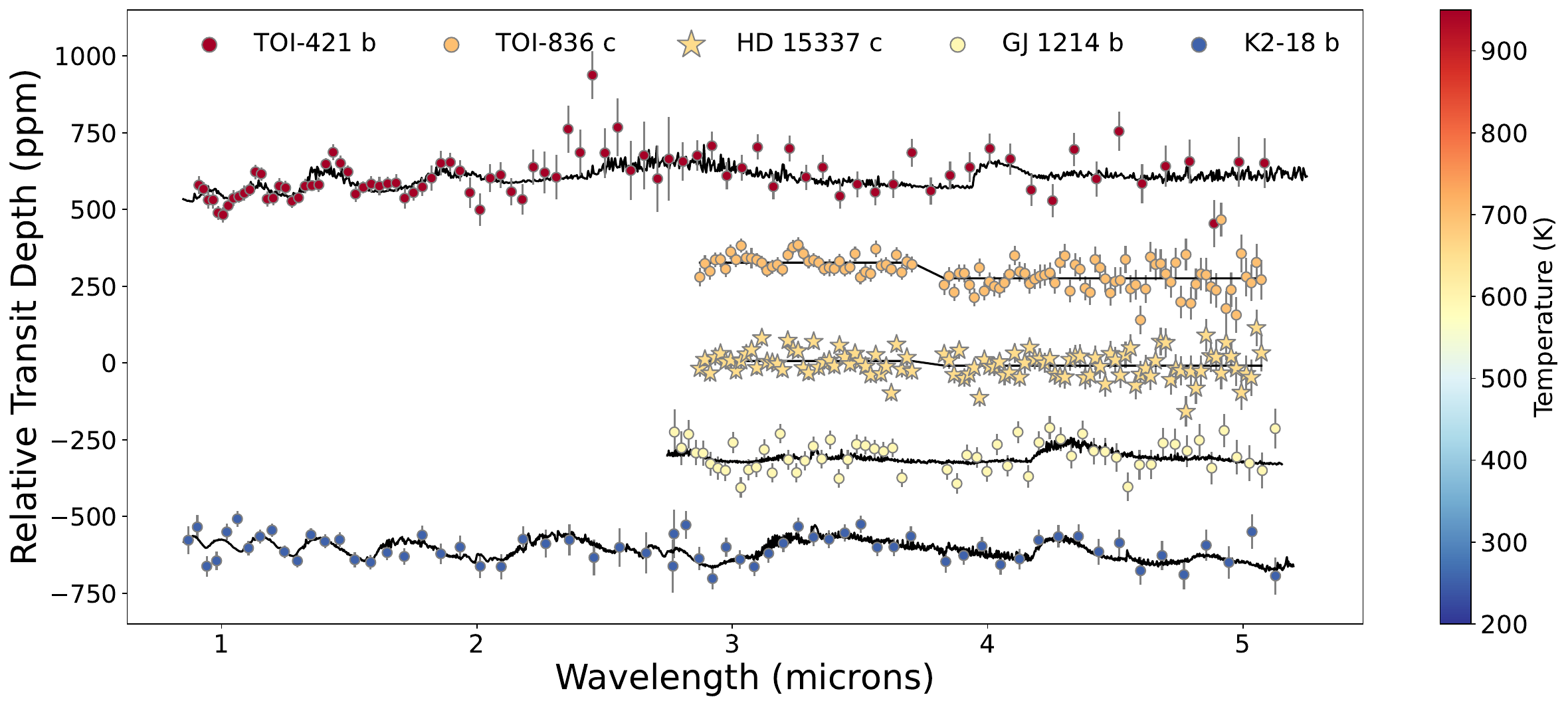} 
\caption{Near-infrared transmission spectra and best-fitting models of the five sub-Neptunes that occupy a similar part of parameter space in Figure~\ref{fig:mr}. The spectrum of TOI-421~b and its accompanying best-fitting model are from \cite{Davenport2025}, the spectrum and best-fitting model for GJ 1214~b are from \cite{Schlawin2024}, and the spectrum and best-fitting model for K2-18~b are from \cite{Madhu2023}. Neither TOI-836~c \citep{Wallack2024} nor HD 15337~c have detected features so we show the corresponding best nonphysical model consisting of a step function to account for the offsets between the two detectors for these planets. There are currently no NIRISS/SOSS observations of GJ 1214~b, HD 15337~c, or TOI-836~c.}
\label{fig:spec}   
\end{centering}
\end{figure*}

HD 15337~c joins a growing population of sub-Neptunes with similar masses ($\sim$6-8 M$_{\oplus}$) and radii ($\sim$2.5-2.6 R$_{\oplus}$; see Figure \ref{fig:mr}) that have been observed with JWST NIRSpec G395H/M and/or NIRISS/SOSS (Figure \ref{fig:spec}). Among these, TOI-836~c \citep{Wallack2024} and GJ 1214~b \citep{Schlawin2024, Ohno2025} are at similar equilibrium temperatures to HD 15337~c (500-600 K), and all three planets show relatively flat spectra from 3-5 $\mu$m indicative of high ($>$300 $\times$ solar) metallicity atmospheres and/or a thick aerosol layer located at pressures $\leq$0.1 mbar. While the composition of these aerosols is currently unknown, their existence at low pressures suggests that they are organic hazes arising from photochemistry, which tends to be most vigorous at lower pressures \citep[e.g.,][]{kawashima2019,Gao2023}. Alternatively, condensate clouds composed of KCl and/or ZnS may also exist at these temperatures \citep{morley2015}, though they would require high atmospheric mixing rates \citep{gao2018gj1214b} and/or high particle porosity \citep{ohno2020agg} to persist at the low pressures we have inferred. In contrast, the much cooler ($\sim$250 K) K2-18~b and the much warmer ($\sim$920 K) TOI-421~b show lower metallicity atmospheres that are devoid of observable aerosols \citep{Madhu2023,Davenport2025, Schmidt2025}. To first order, this fits the trend presented in \cite{Dymont2022} and \cite{Brande2024}, who showed that planets with equilibrium temperatures $\sim$500-600 K are expected to have the flattest transmission spectra due to aerosols, while both cooler and warmer planets should have larger amplitude spectral features potentially due to lower aerosol production rates arising from a lower flux of UV photons and lower abundance of gaseous haze precursor species, respectively. 

While a trend in aerosol production with temperature may exist, the atmospheric metallicity of these objects appears to be much more diverse. For example, although GJ 1214~b is likely to possess a $\geq$1000 $\times$ solar metallicity atmosphere \citep{Kempton2023,Gao2023,Ohno2025}, TOI-421~b's atmosphere is near-solar \citep{Davenport2025}. Meanwhile, TOI-836~c and HD 15337~c's atmospheres could be either high metallicity (or even steam-dominated, as they lie on the 50/50 water/rock composition line; Figure \ref{fig:mr}) or low metallicity with a high altitude aerosol layer \citep{Wallack2024}. K2-18~b's atmospheric metallicity is likely $\leq$100 $\times$ solar but remains uncertain due to different data analysis methods yielding different molecular detections \citep{Madhu2023, Schmidt2025} and a degeneracy between models with a shallow liquid water ocean \citep{Madhu2023} and those with a deep magma ocean \citep{Wogan2024,Shorttle2024}. Adding to this diversity are observations of smaller sub-Neptunes with radii $<$2.4 R$_{\oplus}$ (Figure \ref{fig:mr}), which have shown a steam-dominated but aerosol-free (despite having an equilibrium temperature $\sim$600 K) atmosphere for GJ 9827~d \citep{Piaulet-Ghorayeb2024}; featureless and muted spectra for TOI-776~c \citepalias{Teske2025} and GJ 3090~b \citep{Ahrer2025}, respectively, indicative of high metallicities and/or high altitude aerosol layers, but also signatures of atmospheric escape from Ly$\alpha$ \citep{Loyd2025} and metastable helium observations \citep{Ahrer2025};
and a high metallicity ``miscible envelope'' atmosphere for TOI-270~d \citep{Benneke2024}. The relationship between these smaller worlds and the larger sub-Neptunes requires further study.

The observed variety in sub-Neptune atmospheric composition necessitates further theoretical investigations of sub-Neptune evolution, observational characterization of additional sub-Neptune atmospheres, and more in-depth observations of previously targeted objects. In particular, while HD 15337~c, GJ 1214~b, and TOI-836~c have only been observed by JWST using NIRSpec G395H/M, TOI-421~b and K2-18~b have also been observed using NIRISS/SOSS, which covers multiple CH$_{4}$ and H$_{2}$O features crucial for their detection in these planets' atmospheres \citep{Madhu2023, Davenport2025}. Without the same wavelength coverage and the associated sensitivity to aerosols afforded by the shorter wavelengths accessible to NIRISS/SOSS \citep[e.g.,][]{Wakeford2015} for GJ 1214~b, TOI-836~c, and HD 15337~c, it is difficult to unambiguously determine the composition of these planets, their aerosol opacity, and more generally the primary drivers for atmospheric detectability in sub-Neptunes from transmission spectroscopy\footnote{Although GJ 1214~b has been observed using HST WFC3/G141 \citep{Kreidberg2014}, it does not cover a large enough wavelength range to break the metallicity/aerosol opacity degeneracy.}. While both HD 15337 (J=7.553$\pm$0.019) and TOI-836 (J=7.580$\pm$0.023) previously were too bright to observe with NIRISS/SOSS without at least partial saturation with any more than 2 groups, the newly available (as of Cycle 4) NIRCam Dispersed Hartmann Sensor (DHS) mode and the newly available (as of Cycle 5) additional NIRISS SOSS Multistripe Subarrays may allow for shorter wavelength observations of these relatively bright targets, leading to a more robust determination of their atmospheric composition and aerosol mass loading.

\section{Summary \& Conclusions} 
\label{sec:summary}
We present two transit observations of the sub-Neptune HD 15337~c, taken as part of the COMPASS program. We find that our two observations result in consistent transmission spectra across our two independent reductions and rule out to $>3\sigma$ an atmosphere with mean molecular weight $<$9 g mol$^{-1}$ (metallicity $<$310 $\times$ solar) and opaque pressure level greater than a few mbar assuming thermochemical equilibrium. Our constraints on the atmosphere of this planet are qualitatively similar to those of TOI-836~c, another sub-Neptune observed as part of the COMPASS program. 

We achieve median precisions for visit 1 of $\sim$40 ppm for NRS1 and $\sim$70 ppm for NRS2 and for visit 2 of $\sim$30 ppm for NRS1 and $\sim$54 ppm for NRS2 in 30 pixel spectroscopic bins. The decreased precision in visit 1 can likely be attributed to inadequate pre-transit baseline caused by previously undetected TTVs. Owing to the extreme timing precision of JWST, we detect TTVs for this planet for the first time. After combining the JWST TTVs with previous TESS data to extend the observational baseline, we found a significant long-term TTV signal. JWST's timing precision may reveal unexpected TTVs in additional systems, and by combining JWST transit times with TESS data it is possible to identify trends from previously undetected additional planets using only one or two JWST observations.  

HD 15337~c is the last sub-Neptune observed as part of the COMPASS program. All three COMPASS sub-Neptunes -- TOI-836~c \citep{Wallack2024}, TOI-776~c \citetalias{Teske2025}, and HD 15337~c -- have had featureless transmission spectra best-fit by flat lines with a step function to account for an offset between the detectors. The observations of these planets all rule out clear atmospheres with low mean molecular weight, but further constraints outside of upper limits are difficult, given the absence of detected features. However, additional observations, perhaps using NIRCam DHS or the newly available NIRISS SOSS Multistripe Subarrays, could allow for robust determinations of the compositions of these planets and the breaking of aerosol versus mean molecular weight degeneracies. Acquiring a population of sub-Neptunes with atmospheric constraints would allow for a better understanding of sub-Neptune atmospheric compositions and therefore perhaps a better understanding of their formation and evolution.


\section*{}
The data products associated with this manuscript can be
found at doi:10.5281/zenodo.15360236. This work is based on observations made with the NASA/ESA/CSA James Webb Space Telescope. The data were obtained from the Mikulski Archive for Space Telescopes at the Space Telescope Science Institute, which is operated by the Association of Universities for Research in Astronomy, Inc., under NASA contract NAS 5-03127 for JWST. The specific observations analyzed can be accessed via \dataset[DOI: 10.17909/7we6-0487]{http://dx.doi.org/10.17909/7we6-0487}. These observations are associated with program \#2512. Support for program \#2512 was provided by NASA through a grant from the Space Telescope Science Institute, which is operated by the Association of Universities for Research in Astronomy, Inc., under NASA contract NAS 5-03127. This research has made use of the NASA Exoplanet Archive, which is operated by the California Institute of Technology, under contract with the National Aeronautics and Space Administration under the Exoplanet Exploration Program.

This paper includes data collected by the
TESS mission that are publicly available from the Mikulski
Archive for Space Telescopes (MAST). Support for MAST
for non-HST data is provided by the NASA Office of
Space Science via grant NNX13AC07G and by other
grants and contracts. The specific TESS sectors used in this work can be accessed
via doi:10.17909/5149-1615. We acknowledge the use of public TESS data from pipelines
at the TESS Science Office and at the TESS Science Processing
Operations Center.

This work is funded in part by the Alfred P. Sloan Foundation under grant G202114194.

Support for this work was provided by NASA through grant 80NSSC19K0290 to J.T. and N.L.W. S.E.M. is supported by NASA through the NASA Hubble Fellowship grant HST-HF2-51563 awarded by the Space Telescope Science Institute, which is operated by the Association of Universities for Research in Astronomy, Inc., for NASA, under contract NAS5-26555. AA and NMB acknowledge support from NASA'S Interdisciplinary Consortia for Astrobiology Research (NNH19ZDA001N-ICAR) under award number 80NSSC21K0597.

This work benefited from the 2022 and 2023 Exoplanet Summer Program in the Other Worlds Laboratory (OWL) at the University of California, Santa Cruz, a program funded by the Heising-Simons Foundation. 

Co-Author contributions are as follows: N.L.W. led the data analysis of this study. P.G. led the atmospheric modeling efforts. A.M. provided an additional reduction and analysis of the data. M.G.-M. provided the TTV analysis presented in the study. All authors provided detailed comments and conversations that greatly improved the quality of the manuscript.

\software{\texttt{emcee} \citep{Foreman-Mackey2013}, 
\texttt{Eureka!} \citep{Bell2022},  
\texttt{ExoTiC-LD} \citep{Grant2024}, 
\texttt{Matplotlib} \citep{Hunter2007},  
\texttt{NumPy} \citep{harris2020}, 
\texttt{PandExo} \citep{Batalha2017}, 
\texttt{PLATON} \citep{Zhang2025PLATON},
\texttt{pymultinest} \citep{Buchner2014pymultinest}, 
\texttt{SciPy} \citep{Virtanen2020}, 
\texttt{rebound} \citep{Tamayo2020}, 
\texttt{Tiberius} \citep{Kirk2017, Kirk2021}, 
\texttt{TTVFast} \citep{TTVFast} ,
\texttt{TTVFaster} \citep{TTVFaster}
}
\bibliography{references}{}
\bibliographystyle{aasjournal}

\end{document}